\documentclass[
 reprint,
 amsmath,amssymb,
 aps,
 superscriptaddress,
 nofootinbib,
 floatfix
]{revtex4-2}
\usepackage{graphicx}
\usepackage{dcolumn}
\usepackage{bm}
\newcommand\barparena[1]{\overset{\scriptscriptstyle(-)}{#1}}
\usepackage[colorlinks=true, allcolors=blue]{hyperref}

\begin{document}
\preprint{APS/123-QED}
\title{Dynamic Competition of Fast and Collisional Neutrino Flavor Instabilities \\ with Collisional Damping in Spatially Inhomogeneous Systems}
\author{Shota Takahashi}
\email{tkst1228@g.ecc.u-tokyo.ac.jp}
\affiliation{Department of Physics, The University of Tokyo, 7-3-1 Hongo, Bunkyo, Tokyo 113-0033, Japan}

\author{Hiroki Nagakura}
\affiliation{Division of Science, National Astronomical Observatory of Japan, 2-21-1 Osawa, Mitaka, Tokyo 181-8588, Japan}

\author{Masamichi Zaizen}
\affiliation{Department of Earth Science and Astronomy, The University of Tokyo, Tokyo 153-8902, Japan}

\author{Chinami Kato}
\affiliation{Department of Physics, The University of Tokyo, 7-3-1 Hongo, Bunkyo, Tokyo 113-0033, Japan}

\author{Jiabao Liu}
\affiliation{Department of Physics and Applied Physics, School of Advanced Science and Engineering, Waseda University, Tokyo 169-8555, Japan}

\date{\today}
\begin{abstract}
Neutrino flavor evolution in dense astrophysical environments such as core-collapse supernova (CCSN) is influenced by collective effects. While the Fast Flavor Instability (FFI) and the Collisional Flavor Instability (CFI) are recognized as key drivers of rapid flavor conversion, their non-linear competition with collisional damping in spatially inhomogeneous systems remains poorly understood. Motivated by recent findings that FFI and resonance-like CFI co-occur in the post-bounce phase in CCSN, we scrutinize their dynamic competitions and asymptotic states. To this end, we perform numerical simulations of the quantum kinetic neutrino transport, incorporating both spatial advection and the collision terms. We demonstrate that the interplay between these coexisting neutrino flavor instabilities and collisions leads to rich dynamics. Rather than merely inducing simple decoherence, collisional damping can substantially alter the overall dynamics of collective flavor oscillations, driving the system through complex evolutionary pathways. In all cases where flavor instability develops, we find that the system converges to the same flavor-equilibrated asymptotic state, despite the diversity of intermediate dynamics. Our results suggest that realistic collisional effects drive the system to an asymptotic state distinct from the one predicted by the collisionless FFI picture. This highlights the importance of incorporating collisional effects when modeling the asymptotic outcome of flavor conversion in CCSN models.
\end{abstract}

\maketitle

\section{\label{sec:introduction}Introduction}
Neutrinos play a pivotal role in the dynamics and nucleosynthesis of dense astrophysical environments, such as core-collapse supernovae (CCSNe) and binary neutron star mergers (BNSMs). In the deepest regions of these environments, the neutrino number density is sufficiently high that neutrino-neutrino coherent forward scattering leads to collective neutrino oscillations, altering the flavor evolution \cite{Duan_2010, Mirizzi_2015, Tamborra_2021, Richers_2022, Capozzi_2022, Fischer_2024, Johns_2025}. Understanding these collective effects is essential for accurately modeling the physical processes in compact objects.

Over the past few years, it has become evident that neutrino self-interactions can trigger rapid flavor conversions on timescales orders of magnitude shorter than those of vacuum oscillations. This phenomenon, known as the Fast Flavor Instability (FFI) \cite{Sawyer_2005, Sawyer_2016}, has been the subject of intense theoretical and numerical investigations. It is now well-established that the FFI is triggered by the presence of zero-crossings in the angular distribution of the difference between electron lepton number (ELN) and heavy-lepton number (XLN) \cite{Izaguirre_2017, Morinaga_2021, Dasgupta_2022, Dasgupta_2025}. A vast body of recent literature has focused on diagnosing these crossings in hydrodynamical simulations \cite{Dasgupta_2018,Abbar_2020,Nagakura_2021,Nagakura_2021_3w, Cornelius_2025}, evaluating the exponential growth rates (\cite{Abbar_2019_ffisn, Abbar_2020_ffisn, Nagakura_2019, Delfan_Azari_2020, Morinaga_2020, Glas_2020, Harada_2022, Akaho_2023, Akaho_2024, xiong_2025} for CCSNe and \cite{Wu_2017, Just_2022, Richers_2022_bnsm, Froustey_2024, Froustey_2026, Kawaguchi_2025, Nagakura_2025} for BNSMs) and exploring the highly non-linear asymptotic states of FFI, which often tend toward flavor quasi-equipartition in collisionless regimes (\cite{Wu_2017, Bhattacharyya_2021_FFD, Bhattacharyya_2022, Richers_2021, Richers_2022_codecomparison, Zaizen_2023_quasi, Xiong_2023_asymp} in local calculations; \cite{Nagakura_2023, Nagakura_2023_PRD} in global ones).

In parallel with the developments in FFI, the interplay between fast flavor conversions and incoherent collisions has garnered significant attention. Several earlier studies have investigated the coupling between FFI and collisional effects, exploring the impacts of scattering \cite{Shalgar_2021, Kato_2022, Hansen_2022, Sasaki_2022, Johns_2022, Padilla_Gay_2022, Azari_2024} and emission/absorption \cite{Kato_2023_Emission} in homogeneous setups, as well as in spatially inhomogeneous models \cite{Martin_2021, Sigl_2022}. Recently, substantial efforts have also been made to rigorously incorporate collision terms into global QKE simulations to evaluate their macroscopic consequences \cite{Xiong_2023, Nagakura_2023_PRL, Nagakura_2023_PRD, Xiong_2024_PRD, Shalgar_2024}. Traditionally, these collisional processes were primarily thought to induce quantum decoherence and damp flavor oscillations \cite{Blaschke_2016, Richers_2019}. However, recent breakthroughs have revealed that flavor-dependent collisions can actively drive a distinct class of instabilities known as the Collisional Flavor Instability (CFI) \cite{Johns_2023}, leading to rich phenomenological outcomes such as resonance-like instability enhancements \cite{Xiong_2023_CFI, Liu_2023, Kato_2023_Swap}, muon-induced CFIs \cite{Jiabao_2024_Muon}, and collisional flavor swap/equilibrium \cite{Zaizen_2025_Spectral, Froustey_2025}.

A critical realization, however, is that FFI and CFI do not occur in isolation in realistic astrophysical environments. Recent multi-dimensional Boltzmann simulations of CCSNe have demonstrated that the FFI and the resonance-like CFI can, in general, co-occur in the same spatial regions \cite{Akaho_2024, Nagakura_2025}. This spatial overlap is not a rare exception but rather a general feature of the post-bounce CCSN core. Therefore, instead of conducting a generic parameter survey across all isolated regimes of FFI, CFI, and collisions, understanding the interplay arising from their coexistence is essential. On the one hand, the coexisting instabilities may drive a rapid growth of flavor coherence; on the other hand, collisional terms work to isotropize the angular distributions. How these competing effects shape the non-linear evolution and the resulting asymptotic state remains an open question, with direct implications for modeling flavor transport in CCSNe.

In this paper, we perform non-linear, spatially inhomogeneous neutrino transport simulations by solving quantum kinetic equations (QKEs) that simultaneously incorporate FFI, CFI, and collisional relaxation. Our numerical framework includes spatial advection, allowing us to capture the propagation of neutrinos alongside their collective mixing in a computational domain employing periodic boundary conditions. We note that the spatial inhomogeneity in our setup refers to the neutrino field itself, which develops spatial structure from the initial random perturbations, whereas the background medium is assumed to remain homogeneous, isotropic, and static throughout the evolution. By tracking the long-term evolution driven by the competition between collective instabilities and collisional damping, we aim to uncover how this dynamic interplay dictates the non-linear evolution and whether the system converges to a universal asymptotic state.

This paper is organized as follows. In Sec.~\ref{sec:numerical_setup}, we introduce our numerical setup, including the evolution equations for the neutrino density matrices and the specific implementation of the collisional relaxation terms. In Sec.~\ref{sec:overall_features}, we present the main results of our simulations, focusing on the flavor evolution and the characteristics of the resulting asymptotic states. In Sec.~\ref{sec:analysis}, we delve deeper into the underlying physical dynamics and analyze the detailed temporal evolution of the system. Finally, in Sec.~\ref{sec:discussion}, we summarize our findings and discuss their physical implications, limitations, and astrophysical relevance. Throughout this work, we adopt natural units where $\hbar = c = k_B = 1$.

\section{\label{sec:numerical_setup}Numerical setup}
In this section, we describe the numerical setup and methodology employed in our simulations. Throughout this paper, quantities with an overbar such as $\bar{\rho}$, $\bm{\bar{P}}$, or $\bar{\Gamma}$ denote the corresponding quantities for antineutrinos.

\subsection{Quantum Kinetic Equations}
We perform one-dimensional single-energy ($\varepsilon_\nu = 20\,{\rm MeV}$) numerical simulations of the spatially inhomogeneous quantum kinetic equation (QKE) for neutrinos and antineutrinos, compactly expressed as
\begin{equation}
 i\left(\frac{\partial\barparena{\rho}}{\partial t}+v_z\frac{\partial\barparena{\rho}}{\partial z}\right) = [H,\barparena{\rho}]+i\barparena{C}[\barparena{\rho}],
\end{equation}
where $\barparena{\rho} = \barparena{\rho}(t, z, v_z)$ is the density matrix, $v_z$ is the $z$-component of the velocity, and $\barparena{C}$ is the collision term. Here, we assume axial symmetry in neutrino momentum space around the $z$-axis.

In general, the Hamiltonian $H$ consists of the vacuum oscillation term $H_{\rm vac}$, the matter potential $H_{\rm mat}$, and the neutrino-neutrino self-interaction potential $H_{\nu\nu}$. For antineutrinos, the corresponding Hamiltonian is typically given by $\bar{H} = -H_{\rm vac} + H_{\rm mat} + H_{\nu\nu}$. In this study, however, we neglect the vacuum and matter terms ($H_{\rm vac} = H_{\rm mat} = 0$) because we focus only on FFI and CFI. Consequently, the Hamiltonian is identical for both species ($H = \bar{H}$) and is given solely by the self-interaction term:
\begin{equation}
 H \approx H_{\nu \nu}=\sqrt{2}G_F \int dv_z'(\rho'-\bar{\rho}')(1-v_zv_z'),
\end{equation}
where $G_F$ is the Fermi constant.

In this study, we limit our analysis to a two-flavor neutrino system, consisting of electron neutrinos ($\nu_e$) and heavy-lepton neutrinos ($\nu_x$). Under this assumption, the evolution can equivalently be expressed in terms of polarization vectors:
\begin{equation}
 \label{eq:polarization}
 \begin{aligned}
 \frac{\partial \barparena{P_0}}{\partial t} &= -v_z\frac{\partial \barparena{P_0}}{\partial z}+\barparena{C_0}, \\
 \frac{\partial \barparena{\bm{P}}}{\partial t} &= -v_z\frac{\partial \barparena{\bm{P}}}{\partial z} + \bm{H}_{\nu \nu} \times\barparena{\bm{P}} + \barparena{\bm{C}},
 \end{aligned}
\end{equation}
where the polarization vectors are defined by the relation within the two-flavor framework:
\begin{equation}
 \barparena{\rho} \equiv \frac{1}{2}\big(\barparena{P_0} \bm{I}+\barparena{\bm{P}}\cdot\bm{\sigma}\big).
\end{equation}
The Hamiltonian vector is written as
\begin{equation}
 \bm{H}_{\nu \nu} = \sqrt{2}G_F \int dv_z' \left(\bm{P} - \bm{\bar{P}}\right)(1-v_zv_z').
\end{equation}
Throughout this paper, we define the angle-integrated number density of (anti)neutrinos of flavor $\alpha$ as
\begin{equation}
 \barparena{n_\alpha} \equiv \int_{-1}^{1} dv_z\, \barparena{\rho_{\alpha\alpha}}.
 \label{eq:n_def}
\end{equation}
Under our axisymmetric, single-energy treatment, this convention is consistent with the angular distribution function $g(v_z;\beta)$ introduced below in Sec.~\ref{sec:setup_initial_conditions}, which is normalized as $\int_{-1}^{1} dv_z\, g(v_z;\beta) = 1$, so that $\barparena{n_\alpha^{\rm init}} = \barparena{n_\alpha^{\rm eq}}$ at $t=0$.

\subsection{Collision Term and Background Equilibrium}
In this study, we consider neutrino emission and absorption processes as collisions. Following the prescription in \cite{Blaschke_2016, Richers_2019}, the general collision term is given by
\begin{equation}
 C_{ab} = j_{\nu_a}\delta_{ab} - \left(\langle j\rangle_{ab}+\langle\kappa\rangle_{ab}\right)\rho_{ab},
\end{equation}
where $j_{\nu_a}$ is the emissivity for flavor $a$, $\langle j \rangle_{ab}$ and $\langle \kappa \rangle_{ab}$ are the flavor-averaged emissivity and absorptivity matrices, respectively, and the Pauli-blocking effects are naturally incorporated. The effective collision rate is generally given by $\barparena{\Gamma} = (\barparena{\Gamma_e}+\barparena{\Gamma_x})/2$. In this study, we assume that the charged-current emission and absorption of electron-type neutrinos dominate over the pair processes, and we also neglect all reactions for heavy-lepton neutrinos ($\barparena{\Gamma_x} = 0$) for simplicity. This yields $\barparena{\Gamma} = \barparena{\Gamma_e} / 2$, with $\barparena{\Gamma_e}$ being the electron (anti)neutrino collision rate. Under this assumption, by projecting the matrix collision term onto the $SU(2)$ polarization vector basis, the collision terms for the polarization vectors reduce to
\begin{align}
 \barparena{C_0} &= \barparena{\Gamma}(\barparena{P_0^{\rm eq}}-\barparena{P_0})+\barparena{\Gamma}(\barparena{P_z^{\rm eq}}-\barparena{P_z}), \label{eq:C0} \\
 \barparena{\bm{C}} &=\barparena{\Gamma}(\barparena{\bm{P}^{\rm eq}}-\barparena{\bm{P}})+\barparena{\Gamma}(\barparena{P_0^{\rm eq}}-\barparena{P_0})\bm{\hat{z}}, \label{eq:Cvec}
\end{align}
where $\bm{\hat{z}}$ denotes the unit vector in the $z$-direction of the flavor space. The second terms in both expressions arise from the flavor dependence of the collision rates ($\Gamma_e \neq \Gamma_x = 0$), which couples the evolution of $\barparena{P_0}$ and $\barparena{P_z}$.

One of the key objectives of this study is to systematically investigate how the non-linear phase of flavor conversion depends on the collision rate $\Gamma$. The extent to which the magnitude of $\Gamma$ dictates the evolutionary pathways---such as whether it merely suppresses flavor coherence or fundamentally alters the transition to an asymptotic state---is a central question in this dynamic competition. To explicitly explore this dependence, we run simulations across various collision rates, specifically examining values of $\Gamma \in \{10^{-3}\, \mu, 10^{-4}\, \mu\}$, where $\mu$ is the characteristic neutrino self-interaction potential defined in Sec.~\ref{sec:setup_numerical_methods}. For the antineutrino collision rate, we investigate two distinct cases: a symmetric case ($\bar{\Gamma} = \Gamma$) and an asymmetric case ($\bar{\Gamma} = 0.5\Gamma$).

To evaluate the collision term, we must specify the background thermal equilibrium state, represented by $\barparena{P_0^{\rm eq}}$ and $\barparena{\bm{P}^{\rm eq}}$. The base number densities for this equilibrium state, denoted as $n^{\rm eq}$, are determined according to a Fermi-Dirac distribution with a single energy $\varepsilon_\nu = 20\,{\rm MeV}$, a temperature $T = 6.4\,{\rm MeV}$, and chemical potentials $\mu_{\nu_e} = \mu_{\bar{\nu}_e} = 0$ and $\mu_{\nu_x} = \mu_{\bar{\nu}_x} = -2\,{\rm MeV}$. This yields a heavy-lepton to electron-type equilibrium number density ratio of $n^{\rm eq}_{\nu_x}/n^{\rm eq}_{\nu_e} = n^{\rm eq}_{\bar{\nu}_x}/n^{\rm eq}_{\bar{\nu}_e} = 0.74$ \cite{Kato_2023_Swap}. This parameter set is also designed to trigger the resonance-like CFI. In this background thermal equilibrium state, the angular distribution is perfectly isotropic, which dictates the target state of the collisional relaxation for electron neutrinos. It should be noted that since we neglect the collision terms for heavy-lepton neutrinos ($\barparena{\Gamma_x} = 0$), their thermal equilibrium distributions do not participate in the collisional relaxation during the time evolution; rather, they are used solely to determine the base number densities for the initial conditions.

\subsection{\label{sec:setup_initial_conditions}Anisotropic Initial Conditions}
While the background equilibrium state is completely isotropic, the initial conditions of the simulation are designed to deviate from this isotropy to construct the necessary instability conditions of FFI. We introduce an angular distribution prescribed using the following analytic formula,
\begin{equation}
 g(v_z; \beta) = \frac{1 + \beta v_z}{\int_{-1}^{1}(1 + \beta v_z') dv_z'} = \frac{1}{2}(1 + \beta v_z).
\end{equation}

For the initial condition, $\nu_e$ is distributed isotropically with $\beta = 0$, while all other species share a common parameter $\beta$. This setup is physically motivated to mimic a realistic CCSN environment. In such environments, flavor-dependent decoupling radii naturally yield this angular configuration, because the background matter is still sufficiently dense, $\nu_e$ strongly couples to the matter and remains isotropic. Conversely, $\bar{\nu}_e$, $\nu_x$, and $\bar{\nu}_x$ decouple deeper inside the core, meaning their angular distributions tend to be more forwardly peaked in the region of interest. Consequently, since the angular distributions differ between species, the parameter $\beta$ determines the depth of the initial ELN-XLN crossing.

Based on this anisotropic setup, the initial unperturbed density matrices are constructed by substituting the background equilibrium number densities $n^{\rm eq}$ into the angular distribution function $g(v_z; \beta)$:
\begin{align}
 \rho_{ee}^{\rm init} &= n^{\rm eq}_{\nu_e} g(v_z; 0), \\
 \bar{\rho}_{ee}^{\rm init} &= n^{\rm eq}_{\bar{\nu}_e} g(v_z; \beta), \\
 \rho_{xx}^{\rm init} &= n^{\rm eq}_{\nu_x} g(v_z; \beta), \\
 \bar{\rho}_{xx}^{\rm init} &= n^{\rm eq}_{\bar{\nu}_x} g(v_z; \beta).
\end{align}

Finally, to seed the flavor mixing, we map these diagonal initial density matrices to the polarization vectors and introduce a small off-diagonal spatial perturbation. The initial condition for the polarization vectors is thus given by:
\begin{align}
 \barparena{P_0^{\rm init}} &= \barparena{\rho_{ee}^{\rm init}} + \barparena{\rho_{xx}^{\rm init}}, \\
 \barparena{\bm{P}^{\rm init}} &= \begin{pmatrix}
 (\barparena{\rho_{ee}^{\rm init}} - \barparena{\rho_{xx}^{\rm init}})\epsilon(z) \\
 0 \\
 (\barparena{\rho_{ee}^{\rm init}} - \barparena{\rho_{xx}^{\rm init}})\sqrt{1-\epsilon^2(z)}
 \end{pmatrix},
\end{align}
where $\epsilon(z)$ is a random perturbation within an amplitude of $10^{-6}$.

\subsection{\label{sec:setup_numerical_methods}Numerical Methods}
To simulate the QKE in the fast flavor conversion regime, we numerically solve the evolution equations for the polarization vectors. We adopt the neutrino self-interaction potential $\mu=\sqrt{2}G_F n^{\rm eq}_{\nu_e}$ at the initial condition as the reference scale, measuring both spatial and temporal coordinates in units of $\mu^{-1}$. The computational domain consists of a one-dimensional box of size $L_z = 1000 \, \mu^{-1}$, uniformly resolved by $N_z = 10000$ grid cells under periodic boundary conditions. The momentum phase space is discretized using a 128-point Gauss--Legendre quadrature for the angular distribution. We have carefully confirmed that this combination of spatial and angular resolutions is fine enough to resolve all structures generated throughout the entire simulation without suffering from numerical artifacts \cite{Nagakura_2025_resolution}.

For the numerical integration, we apply a fifth-order weighted essentially non-oscillatory (WENO) scheme to compute the spatial advection terms in Eq.~(\ref{eq:polarization}). The time evolution of the polarization vectors is tracked via a fourth-order strong stability preserving Runge--Kutta scheme with five stages (SSP-RK(5,4)), where the time step is constrained by a Courant--Friedrichs--Lewy (CFL) parameter of $C_{\rm CFL} = 0.4$. Our numerical simulations are performed using the GPU-accelerated quantum kinetic neutrino transport code GANTS-QK; further details of the code implementation can be found in the appendix of \cite{Zaizen_2026}.

\begin{figure*}
    \centering
    \includegraphics[width=1\linewidth]{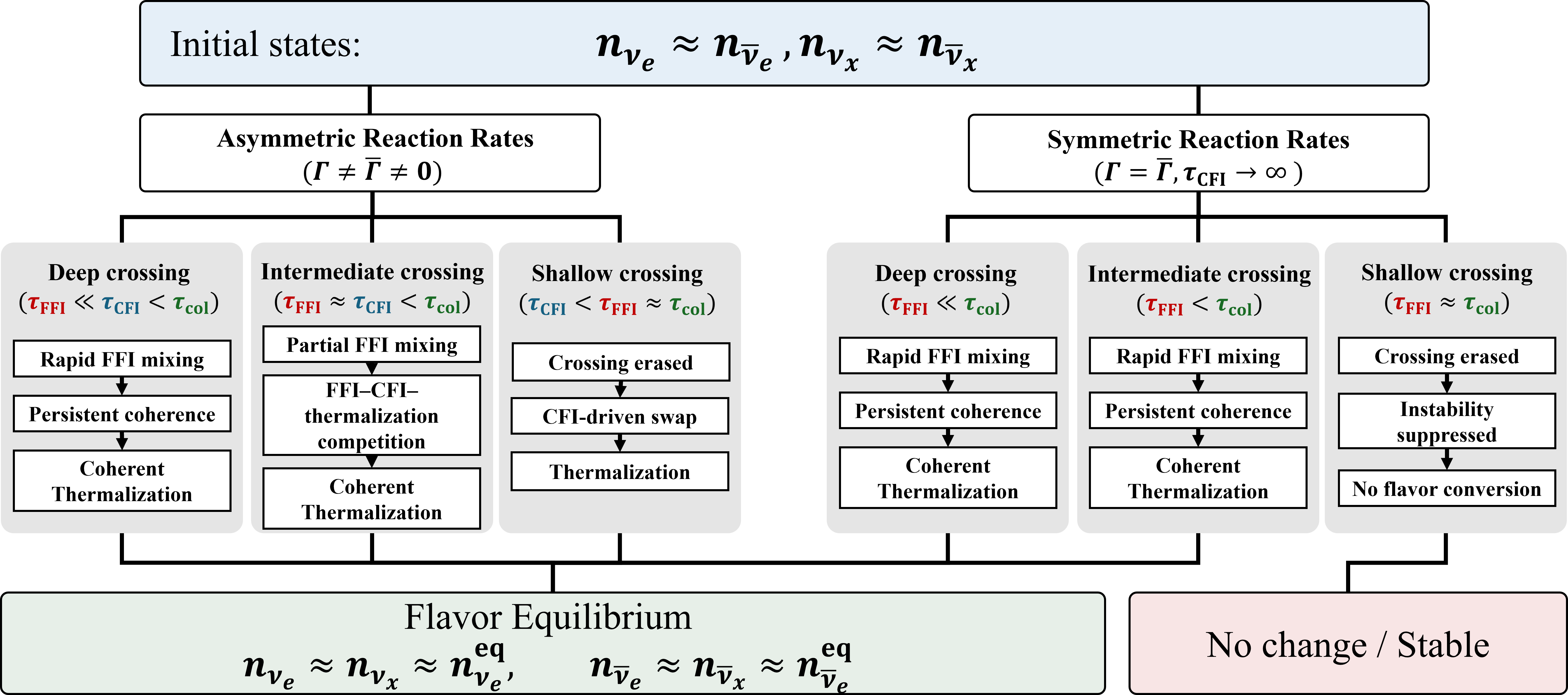}
    \caption{Chart illustrating the evolutionary pathways of neutrino flavor mixing and collisional thermalization. The evolution branches based on two primary conditions: the symmetry of the collision rates (symmetric vs. asymmetric) and the depth of the initial crossing (deep, intermediate, and shallow). The crossing depth regimes are classified by the hierarchy of the relevant timescales: FFI ($\tau_{\text{FFI}}$), CFI ($\tau_{\text{CFI}}$), and collisions ($\tau_{\text{col}}$). The sequential boxes in each column indicate the dominant physical processes at different stages of the evolution. The bottom panels show the final asymptotic state of the system, indicating whether it converges to a universal flavor equilibrium or remains in a stable state with no flavor conversion.}
    \label{fig:summary}
\end{figure*}

\section{\label{sec:overall_features}Overall Features}
In this section, we present the numerical results of the spatially inhomogeneous quantum kinetic neutrino transport. We begin with an overview of the overall trends before describing the detailed dynamics of each model.

\subsection{\label{sec:overview}Overview of Evolutionary Pathways}
The diverse evolutionary pathways observed in our simulations are summarized in the chart shown in Fig.~\ref{fig:summary}. The dynamical evolution of the neutrino gas can be characterized by the interplay between two factors: the ELN-XLN angular crossing depth (parameterized by $\beta$) and the collision rates ($\Gamma$, $\bar{\Gamma}$). This is because these physical parameters directly determine the relative balance among three critical timescales: the growth timescale of the fast flavor instability ($\tau_{\rm FFI}$), that of the collisional flavor instability ($\tau_{\rm CFI}$), and the collisional thermalization timescale ($\tau_{\rm col}$). Ultimately, the dynamic competition among these timescales dictates the non-linear fate of the flavor conversion.

Here, we provide approximate estimates of the growth timescales of both FFI and CFI. For $\tau_{\rm FFI}$, we adopt an approximate method proposed by \cite{Nagakura_2019, Morinaga_2020}, yielding
\begin{equation}
 \begin{aligned}
 \tau_{\rm FFI} = 2 \pi \left| \left(\int_{G_{v}>0} d\Gamma\, G_{v}\right)\left(\int_{G_{v}<0} d\Gamma\, G_{v}\right) \right|^{-1/2},
 \end{aligned}
 \label{eq:approxiGrowth_NaMo}
\end{equation}
where, under our single-energy treatment,
\begin{equation}
 \begin{split}
 d \Gamma_{v} &\equiv dv_z, \\
 G_{v} &\equiv \frac{1}{2}(\langle \rho_{ee} \rangle - \langle \bar{\rho}_{ee} \rangle - \langle \rho_{xx} \rangle + \langle \bar{\rho}_{xx} \rangle).
 \end{split}
 \label{eq:Gamma_and_Gv}
\end{equation}
For $\tau_{\rm CFI}$, we employ an approximate dispersion relation for homogeneous CFI proposed by \cite{Liu_2023}\footnote{The isotropy-breaking branch is not considered here, since its growth rate is comparable to that of the isotropy-preserving branch.},
\begin{equation}
 \omega_{\pm} = -A - i\gamma \pm \sqrt{A^2 - \alpha^2 + i2G\alpha},
 \label{eq:dispersions}
\end{equation}
where the parameters are defined using the spatially averaged number densities as
\begin{equation}
 \begin{aligned}
 A &= \frac{1}{2} (\langle n_{\nu_e} \rangle - \langle n_{\nu_x} \rangle - \langle n_{\bar{\nu}_e} \rangle + \langle n_{\bar{\nu}_x} \rangle), \\
 G &= \frac{1}{2} (\langle n_{\nu_e} \rangle - \langle n_{\nu_x} \rangle + \langle n_{\bar{\nu}_e} \rangle - \langle n_{\bar{\nu}_x} \rangle), \\
 \gamma &= \frac{\Gamma + \bar{\Gamma}}{2}, \\
 \alpha &= \frac{\Gamma - \bar{\Gamma}}{2}.
 \end{aligned}
 \label{eq:CFI_primitive}
\end{equation}
The maximum CFI growth rate, $\mathrm{Im}(\omega_\pm) \equiv \max\bigl(\mathrm{Im}(\omega_+), \mathrm{Im}(\omega_-)\bigr)$, admits the following approximate forms \cite{Liu_2023}:
\begin{equation}
 \mathrm{Im}(\omega_{\pm}) \approx \begin{cases}
 -\gamma + |G\alpha|/|A|, & (A^2 \ge |G\alpha|), \\
 -\gamma + \sqrt{|G\alpha|}, & (A^2 < |G\alpha|),
 \end{cases}
 \label{eq:CFIgrowthrate_isopre}
\end{equation}
corresponding to the non-resonant and resonance-like limits, respectively. We then estimate the CFI time scale as
\begin{equation}
 \tau_{\rm CFI} = \frac{2 \pi} {\mathrm{Im}(\omega_{\pm})}.
 \label{eq:tau_CFI}
\end{equation}

To systematically understand the competition among FFI, CFI, and collisional damping/thermalization, we investigate two distinct scenarios: symmetric ($\Gamma = \bar{\Gamma}$) and asymmetric ($\Gamma \neq \bar{\Gamma}$) collision rates, specifically to switch the CFI on or off. We note that symmetric rates are unrealistic in neutron-rich environments as in CCSNe, where excess neutrons enhance $\nu_e$ absorption over $\bar{\nu}_e$ absorption ($\Gamma > \bar{\Gamma}$). This idealized baseline is, however, useful to isolate the pure interplay between the FFI and decoherent collisional thermalization. In contrast, the asymmetric setup ($\bar{\Gamma} = 0.5\Gamma$) incorporates this more realistic relationship between the collision rates, enabling us to explore the more complex dynamics among the FFI, CFI, and thermalization.

We find that the dynamical outcomes are categorized into three distinct regimes---labeled as ``deep,'' ``intermediate,'' and ``shallow'' ELN angular crossing in the third row of Fig.~\ref{fig:summary}---based on the hierarchy among these three timescales. It should be noted that these terms do not refer to the absolute geometrical depth of the initial angular crossing. Rather, they describe the strength of the crossing \textit{relative} to the degree of thermalization. Because the classification is determined by the full timescale hierarchy rather than by $\beta$ alone, the same value of $\beta$ may fall into different regimes depending on the collision rates ($\Gamma$, $\bar{\Gamma}$): \begin{itemize}
    \item \textbf{Deep crossing} ($\tau_{\rm FFI} (\ll \tau_{\rm CFI}) \ll \tau_{\rm col}$): The FFI growth is sufficiently fast, completely overcoming CFI and the collisional damping.
    \item \textbf{Intermediate crossing} ($\tau_{\rm FFI} (\approx \tau_{\rm CFI}) < \tau_{\rm col}$): The instability is faster than the thermalization, but collisions are strong enough to alter the dynamics as well as subsequent asymptotic behaviors.
    \item \textbf{Shallow crossing} ($(\tau_{\rm CFI} <) \ \tau_{\rm FFI} \approx \tau_{\rm col}$): The instability timescale is comparable to the thermalization timescale, allowing frequent collisions to heavily suppress the flavor instabilities before they can fully develop.
\end{itemize}

As depicted at the bottom of Fig.~\ref{fig:summary}, in all cases where flavor instability develops, the number densities of the different flavors converge asymptotically to the equilibrium values of the electron-type species, specifically $n_{\nu_e} \approx n_{\nu_x} \approx n_{\nu_e}^{\rm eq}$ and $n_{\bar{\nu}_e} \approx n_{\bar{\nu}_x} \approx n_{\bar{\nu}_e}^{\rm eq}$, despite the highly diverse evolutionary pathways. The sole exception is the shallow crossing in the symmetric baseline, where the instability is entirely suppressed and no flavor conversion occurs.

With this classification and the ultimate asymptotic state in mind, we will overview the macroscopic evolution of the symmetric baseline in Sec.~\ref{sec:sym}, followed by the asymmetric setup in Sec.~\ref{sec:asym}. Because this section strictly serves as an overview of all possible scenarios, a thorough investigation of the underlying physical phenomena for each dynamical feature will be deferred to Sec.~\ref{sec:analysis}.

\subsection{\label{sec:sym}Symmetric Collision Rates ($\Gamma = \bar{\Gamma}$)}
\begin{figure*}
    \centering
    \includegraphics[width=0.75\linewidth]{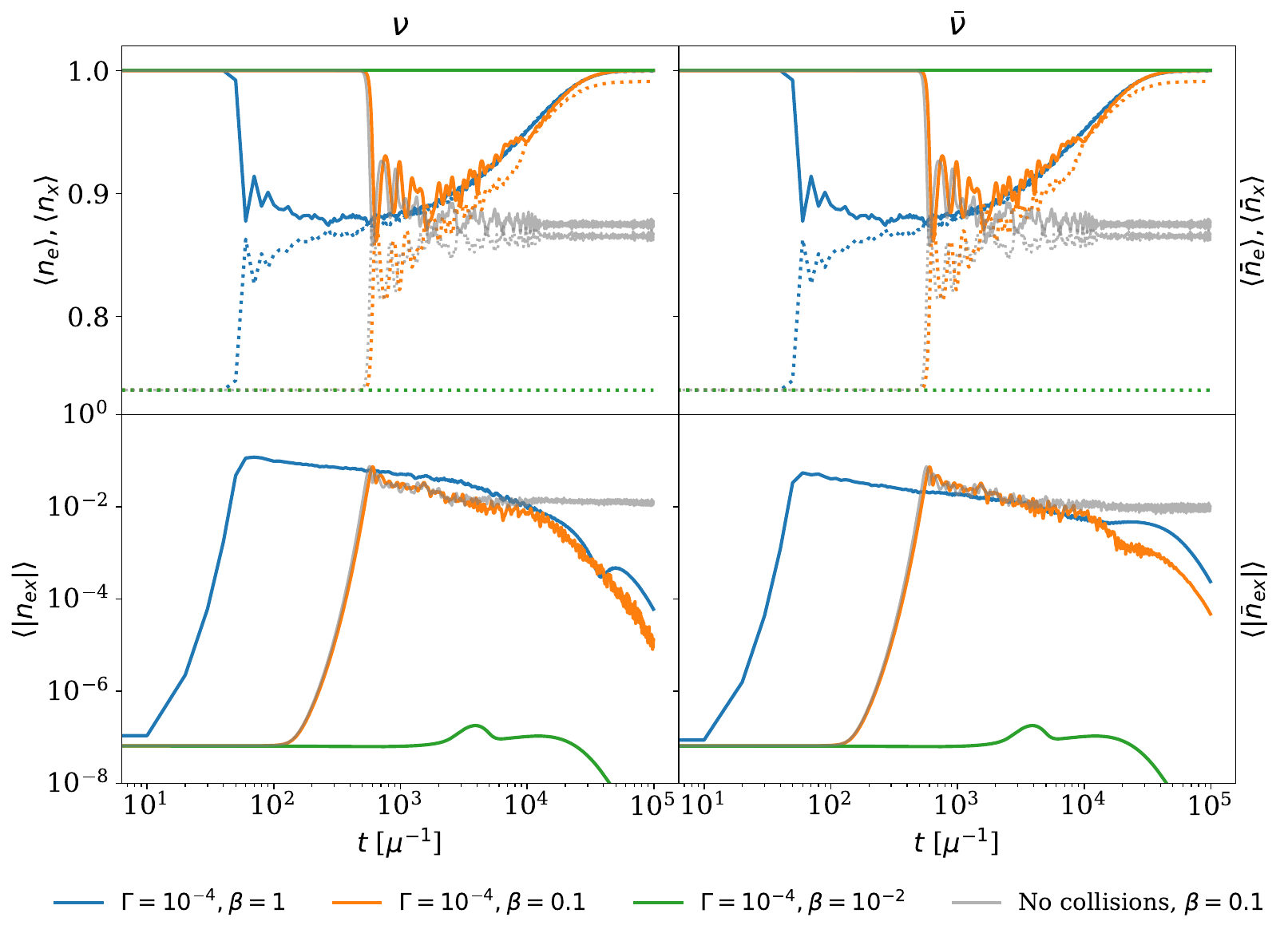}
    \caption{Time evolution of the spatially-averaged angular moments in the idealized case of symmetric collision rates ($\bar{\Gamma} = \Gamma$) for a fixed collision rate $\Gamma = 10^{-4}$. The left and right columns show neutrinos and antineutrinos, respectively. Upper panels display the diagonal components $\langle n_{\nu_e} \rangle$ (solid) and $\langle n_{\nu_x} \rangle$ (dotted), while lower panels show the absolute off-diagonal component $\langle |n_{\mathrm{ex}}| \rangle$. Different colors represent initial crossing depths: $\beta = 1.0$ (Deep crossing, blue), $\beta = 0.1$ (Intermediate crossing, orange), and $\beta = 0.01$ (Shallow crossing, green). As a reference, the result with $\beta = 0.1$ and $\Gamma = 0$ is displayed as gray lines.}
    \label{fig:symmetric_low_gamma}
\end{figure*}

We begin by analyzing the symmetric baseline ($\Gamma = \bar{\Gamma}$), which strictly prohibits the CFI ($\tau_{\text{CFI}} \to \infty$) and isolates the pure competition between the FFI and collisions. Figure~\ref{fig:symmetric_low_gamma} displays the results for a fixed representative collision rate of $\Gamma = 10^{-4}$ with varying relative crossing depths ($\beta$).

\subsubsection{Deep Crossing}
In this regime ($\beta = 1.0$, blue lines in Fig.~\ref{fig:symmetric_low_gamma}), the FFI dominates the early dynamics. Because the initial state contains an excess of electron flavors compared to heavy-lepton flavors, the FFI drives the number densities toward almost perfect flavor equipartition, which significantly reduces the number of electron-type neutrinos.

Following this initial mixing, the system transitions into a persistent coherent state. During the subsequent thermalization phase, collisional processes---which act exclusively on $\nu_e$ and $\bar{\nu}_e$---work to increase these depleted densities toward their thermal equilibrium values. However, because the system maintains strong flavor coherence, continuous flavor conversion immediately distributes this newly generated electron-type population to $\nu_x$ and $\bar{\nu}_x$. This specific state---where strong flavor coherence is maintained throughout the thermalization phase---closely parallels the ``edge of instability'' \cite{Fiorillo_2024} and quasi-steady evolution of flavor conversion reported in \cite{Liu_2024_quasisteady, Urquilla_2025}, wherein the system continuously hovers near marginal stability.In contrast to the previous studies \cite{Fiorillo_2024, Liu_2024_quasisteady}, our simulations include both emission and absorption processes. Our result indicates that the persistent flavor conversion may also arise in such realistic CCSN-like settings.

At later times, the heavy-lepton neutrinos also approach isotropic distributions despite not being directly affected by collisions. Through coherent flavor coupling, they are driven toward the same equilibrium state as the electron-type species, ultimately acquiring the electron-type equilibrium number density rather than their own and leading the system to a universal flavor equilibrium.

\begin{figure*}
    \centering 
    \includegraphics[width=0.75\linewidth]{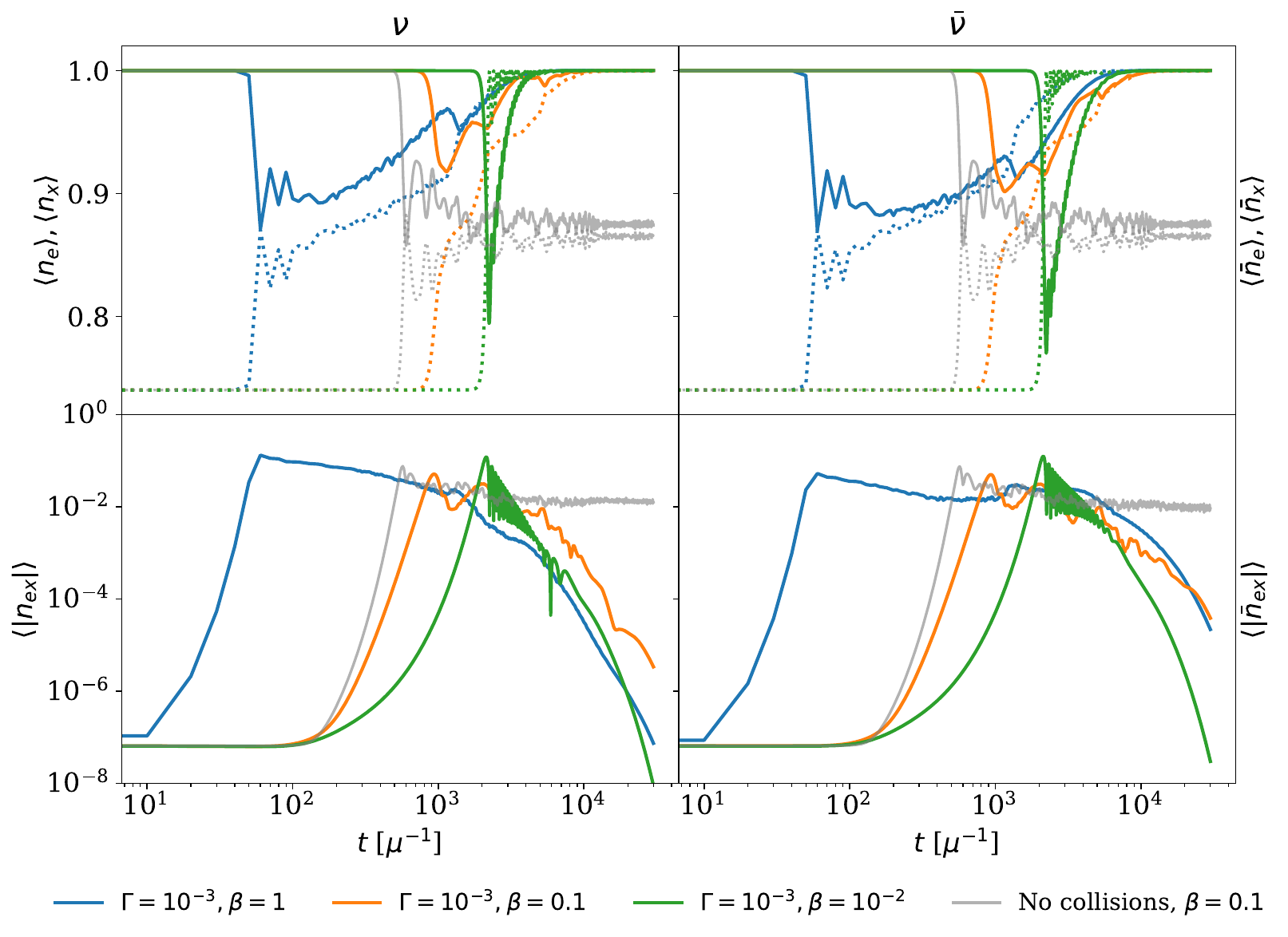} 
    \caption{Same as Fig.~\ref{fig:symmetric_low_gamma}, but for asymmetric collision rates ($\bar{\Gamma} = 0.5\Gamma$) with a higher collision rate of $\Gamma = 10^{-3}$.} 
    \label{fig:asymmetric_high_gamma}
\end{figure*}

\subsubsection{Intermediate Crossing}
For a moderate initial crossing ($\beta = 0.1$, orange lines in Fig.~\ref{fig:symmetric_low_gamma}), the early-phase FFI is weaker compared to the deep crossing case. Consequently, the initial flavor equipartition is also weaker, leaving the system in a state of incomplete mixing. The system then enters a quasi-steady evolution. Similar to the case with deep crossing, the collision plays a pivotal role during the late phase evolution, which can be seen by comparing to gray lines (the case with no collisions) in Fig.~\ref{fig:symmetric_low_gamma}. We also find that a rapid flavor mixing between $\nu_e$ and $\nu_x$ emerges around $t \sim 10^4$. The physical mechanism driving this rapid mixing is the collisional modification of the neutrino angular distributions, which alters the FFI eigenmode structure and is discussed in detail in Sec.~\ref{sec:analysis_modification}.

It is worth noting that the asymptotic states of both $n_{\nu_x}$ and $n_{\bar{\nu}_x}$ are slightly lower than those in electron-type neutrinos. Because the FFI is not as dominant in this model, collisions disrupt flavor coherence before perfect equilibration can be achieved, leaving behind the heavy-lepton neutrino and antineutrino densities saturated slightly below their exact equilibrium levels. Nevertheless, this discrepancy is quantitatively negligible, and the system can be safely regarded as having effectively reached a universal flavor equilibrium.

\subsubsection{Shallow Crossing}
In the shallow regime ($\beta = 0.01$, green lines in Fig.~\ref{fig:symmetric_low_gamma}), collisions are sufficiently frequent to rapidly isotropize the angular distributions, effectively erasing the initial ELN-XLN crossing. Because the CFI is prohibited in this symmetric setup, no flavor conversions can develop once the ELN-XLN crossing condition vanishes. Consequently, the flavor instability is entirely suppressed. The electron-type neutrinos and antineutrinos relax toward their thermal equilibria, whereas the heavy-lepton species, unaffected by collisions, retain their initial distributions.

\subsection{\label{sec:asym}Asymmetric Collision Rates ($\Gamma \neq \bar{\Gamma}$)}
\begin{figure*} 
    \centering 
    \includegraphics[width=0.75\linewidth]{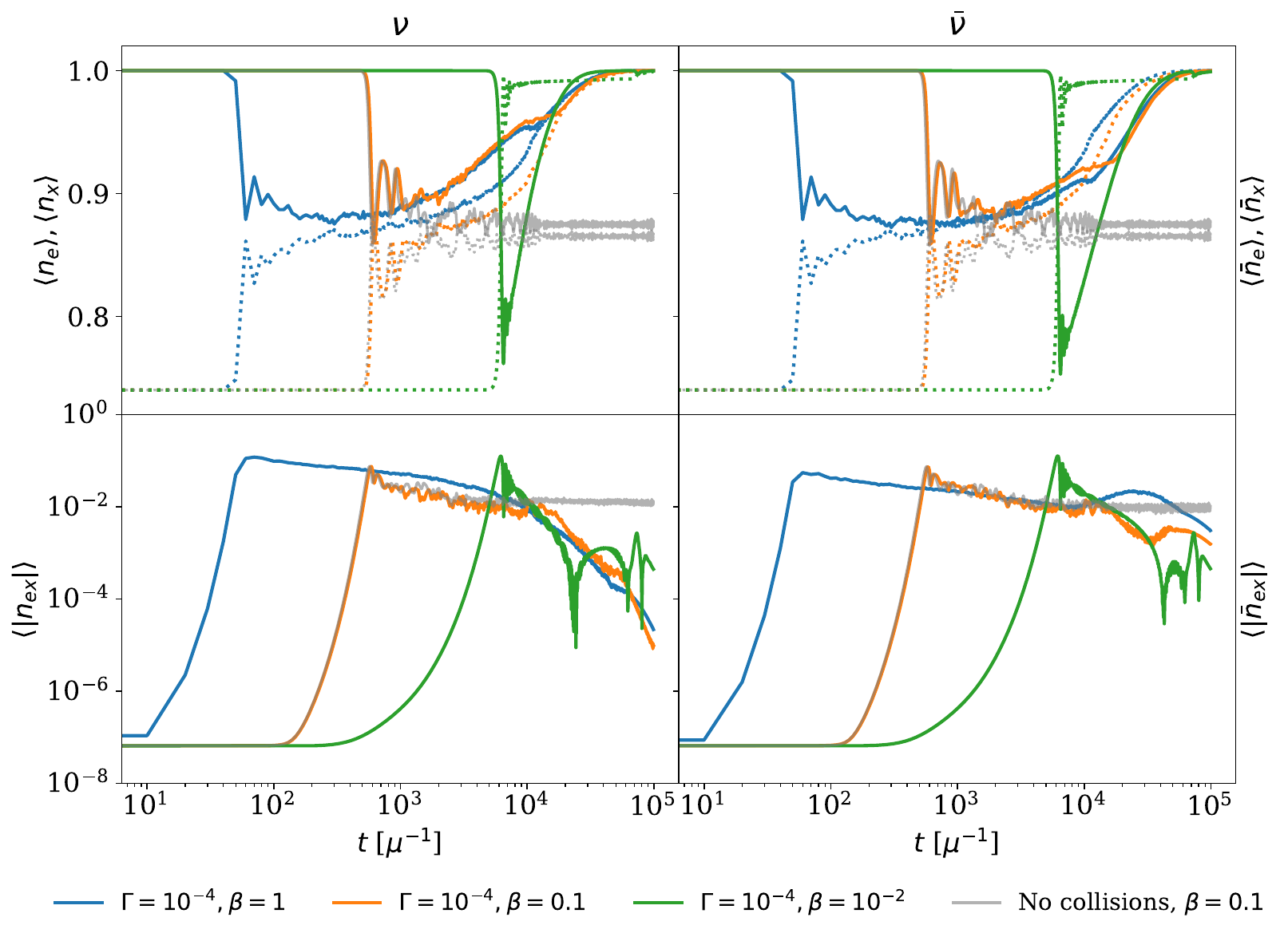} 
    \caption{Same as Fig.~\ref{fig:asymmetric_high_gamma}, but for a lower collision rate of $\Gamma = 10^{-4}$.} 
    \label{fig:asymmetric_low_gamma}
\end{figure*}

Having established the symmetric collision rate case, we now consider cases with asymmetric collision rates ($\bar{\Gamma} = 0.5\Gamma$), as expected in the CCSN environment. In this regime, the collision rate asymmetry enables the system to satisfy the condition for the Collisional Flavor Instability (CFI). This introduces a finite $\tau_{\text{CFI}}$ that actively competes with both the FFI and collisional decoherence. To investigate this interplay, we examine a high collision rate ($\Gamma = 10^{-3}$, Fig.~\ref{fig:asymmetric_high_gamma}) and a low collision rate ($\Gamma = 10^{-4}$, Fig.~\ref{fig:asymmetric_low_gamma}). As already outlined in Sec.~\ref{sec:overview}, all cases in this regime ultimately converge to the identical universal flavor equilibrium.

\subsubsection{Deep Crossing}
This regime encompasses three cases: the large initial crossings ($\beta = 1.0$) for both collision rates ($\Gamma = 10^{-3}$ and $10^{-4}$, blue lines in Figs.~\ref{fig:asymmetric_high_gamma} and \ref{fig:asymmetric_low_gamma}), as well as the moderate crossing ($\beta = 0.1$) for the collision rate of $\Gamma = 10^{-4}$ (orange lines in Fig.~\ref{fig:asymmetric_low_gamma}). It should be noted that the $\beta = 0.1$ case with $\Gamma = 10^{-4}$ is classified as ``deep'' here, in contrast to its ``intermediate'' classification in the symmetric baseline. As discussed in Sec.~\ref{sec:overview}, the regime labels are determined by the full timescale hierarchy, not by $\beta$ alone. With the lower collision rate $\Gamma = 10^{-4}$, both $\tau_{\rm CFI}$ and $\tau_{\rm col}$ are sufficiently long that the FFI remains the dominant instability ($\tau_{\rm FFI} \ll \tau_{\rm CFI} < \tau_{\rm col}$), placing this case in the deep crossing regime.

In all three cases, the FFI drives rapid initial flavor conversion. However, unlike the symmetric deep crossing case where near-complete equipartition is achieved, the system here exhibits incomplete mixing---a behavior reminiscent of the intermediate crossing of the symmetric baseline. This is because the asymmetric collision rates ($\Gamma > \bar{\Gamma}$) drive the ELN angular distribution upward toward its isotropic equilibrium value more rapidly than in the symmetric case, suppressing the FFI more efficiently before it can achieve complete equipartition and leaving the system in a quasi-steady coherent state.

Following this incomplete mixing, the system enters a quasi-steady evolution phase. Just as in the intermediate crossing of the symmetric baseline, collisions gradually modify the ELN angular distribution. Because the collision rates are asymmetric ($\Gamma > \bar{\Gamma}$), $\nu_e$ thermalize more rapidly than their antiparticles, leading to differences in the evolution of their angular distributions. Nevertheless, the overall thermalization process remains qualitatively similar to that in the symmetric case.

At a later stage, we also find that this collisional modification triggers a sudden mixing where $\nu_e$ and $\nu_x$ reach near-equipartition. The detailed mechanism of this transition is discussed in Sec.~\ref{sec:analysis_modification}. Despite the different late phase evolution from symmetric collision rate models, we find that the system undergoes coherent equilibration and successfully reaches the universal flavor equilibrium.

\subsubsection{Intermediate Crossing}
When the initial crossing is moderate and the collision rate is relatively high ($\beta = 0.1$ with $\Gamma = 10^{-3}$, orange lines in Fig.~\ref{fig:asymmetric_high_gamma}), the system falls into the intermediate crossing regime of the asymmetric case, where the growth timescales of the FFI and CFI become comparable. It should be noted, however, that this model has distinct features from the intermediate crossing of the symmetric baseline. Because the growth rates of the FFI and CFI are comparable, their competition disrupts the formation of a quasi-steady coherent state. Instead, the system evolves through an entirely distinct dynamical path in which all three processes---FFI, CFI, and collisional thermalization---coevolve to drive multiple large-amplitude mixing events. We defer the detailed discussion to Sec.~\ref{sec:analysis_competition}. Despite taking this highly dynamic evolutionary path, the system ultimately converges to the same universal flavor equilibrium. This result underscores the robustness of the final asymptotic state.

\subsubsection{Shallow Crossing}
\label{subsec:asymShallowCross}
The asymmetric collision model with shallow initial crossing ($\beta = 0.01$, green lines in Figs.~\ref{fig:asymmetric_high_gamma} and \ref{fig:asymmetric_low_gamma}) exhibits the most pronounced deviations from the symmetric cases. In the early phase, the initial ELN-XLN crossing is rapidly eliminated by frequent collisions, effectively suppressing the FFI. However, due to the asymmetric collision rates under $G>0$ and $A=0$ [see Eqs.~(\ref{eq:CFI_primitive}) and (\ref{eq:CFIgrowthrate_isopre}) for CFI], the system triggers a resonance-like CFI, which becomes the dominant driver of the dynamics. Unlike the symmetric case, which remained completely stable, this CFI induces a rapid and extreme flavor conversion in the nonlinear phase, known as flavor swap. This can be seen in the green lines in the upper panels of Figs.~\ref{fig:asymmetric_high_gamma} and \ref{fig:asymmetric_low_gamma}, where the sudden flavor swap occurs around $t \sim 3000$ for $\Gamma = 10^{-3}$ and $t \sim 10^4$ for $\Gamma = 10^{-4}$. During the phase with strong flavor conversion, however, collisional thermalization gradually hampers flavor conversion, causing an incomplete swap. Nevertheless, the large flavor coherence can sustain flavor mixing during the collisional thermalization. Consequently, the system asymptotically approaches a universal flavor equilibrium.

It is worth emphasizing that our result demonstrates the robustness of the flavor swap driven by the resonance-like CFI even in spatially inhomogeneous environments. This suggests that collisional flavor swap originally identified in homogeneous QKE simulations \cite{Kato_2023_Swap} can occur in more realistic environments.

\section{\label{sec:analysis}Detailed Analysis of Dynamical Features}
As demonstrated in Sec.~\ref{sec:overall_features}, except for the stable scenario where the FFI is completely suppressed (the symmetric shallow-crossing case), the asymptotic states of the neutrino gas are found to be nearly universal across different setups. This universality arises because both the thermalization and decoherence timescales are governed by the collision rate, operating on the order of $\Gamma^{-1}$ timescales longer than those of flavor conversions. Consequently, as long as the system undergoes flavor mixing and evolves over this characteristic timescale, it can settle into a nearly identical flavor-equilibrated state.

Although this asymptotic state is predictable for all unstable cases, the evolutionary paths towards the asymptotic state exhibit rich diversity across different models. The intermediate dynamics are, indeed, very sensitive to the physical parameters, such as the initial crossing depth ($\beta$) and the symmetry of the collision rates ($\Gamma$ vs. $\bar{\Gamma}$). Because the nonlinear interplay between flavor conversion and collisional damping is strongly parameter-dependent, the system can pass through various distinct transient phases before reaching equilibrium. To elucidate the physics behind this diversity, the following subsections delve into the underlying mechanisms driving some intriguing phenomena observed in our models.

\subsection{\label{sec:analysis_modification}Collisional Modification of Angular Distributions and FFI Eigenmode Transition}

In several of our models, we observe that, following the initial fast flavor conversion and the quasi-steady phase, a sudden equalization of $\langle n_{\nu_e} \rangle$ and $\langle n_{\nu_x} \rangle$ occurs. Specifically, this phenomenon is seen in the symmetric intermediate case with $(\Gamma, \beta) = (10^{-4}, 0.1)$ (orange lines in Fig.~\ref{fig:symmetric_low_gamma}, around $t \sim 10^4$), as well as the asymmetric deep-crossing cases with $(\Gamma, \beta) = (10^{-3}, 1.0)$ (blue lines in Fig.~\ref{fig:asymmetric_high_gamma}, around $t \sim 10^3$), $(\Gamma, \beta) = (10^{-4}, 1.0)$ (blue lines in Fig.~\ref{fig:asymmetric_low_gamma}, around $t \sim 10^4$), and $(\Gamma, \beta) = (10^{-4}, 0.1)$ (orange lines in Fig.~\ref{fig:asymmetric_low_gamma}, around $t \sim 10^4$). Since both the FFI and CFI could in principle drive such secondary flavor mixing, the origin of this phenomenon is not immediately obvious. However, we find that the CFI growth rate $\mathrm{Im}(\omega_\pm)$ [defined in Eq.~(\ref{eq:CFIgrowthrate_isopre})] remains negative when this sudden mixing occurs, indicating that the CFI is not responsible.

\begin{figure}
    \centering
    \includegraphics[width=1\linewidth]{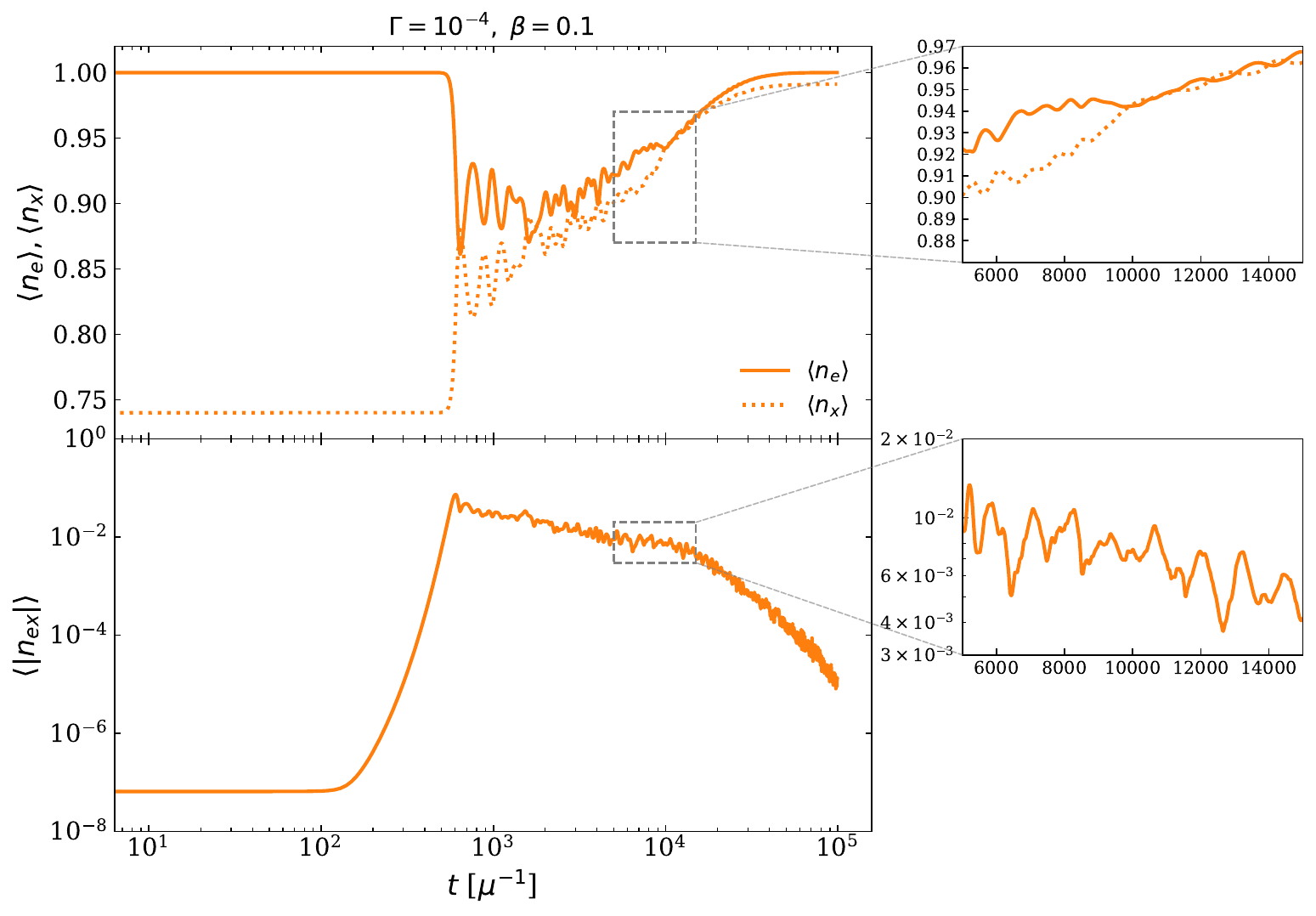}
    \caption{Time evolution of the spatially-averaged angular moments; the diagonal components $\langle n_{\nu_e} \rangle$ (solid) and $\langle n_{\nu_x} \rangle$ (dotted) (upper panel) and the off-diagonal component $\langle |n_{\mathrm{ex}}| \rangle$ (lower panel) for the symmetric case with $(\Gamma, \beta) = (10^{-4}, 0.1)$. The left panels show the full time range on a logarithmic scale, while the right panels focus on the collision-induced sudden mixing event for $t \in [5000, 15000]$ on a linear scale. The dashed rectangles indicate the zoomed region.}
    \label{fig:magnify_sym}
\end{figure}

\begin{figure}
    \centering
    \includegraphics[width=1\linewidth]{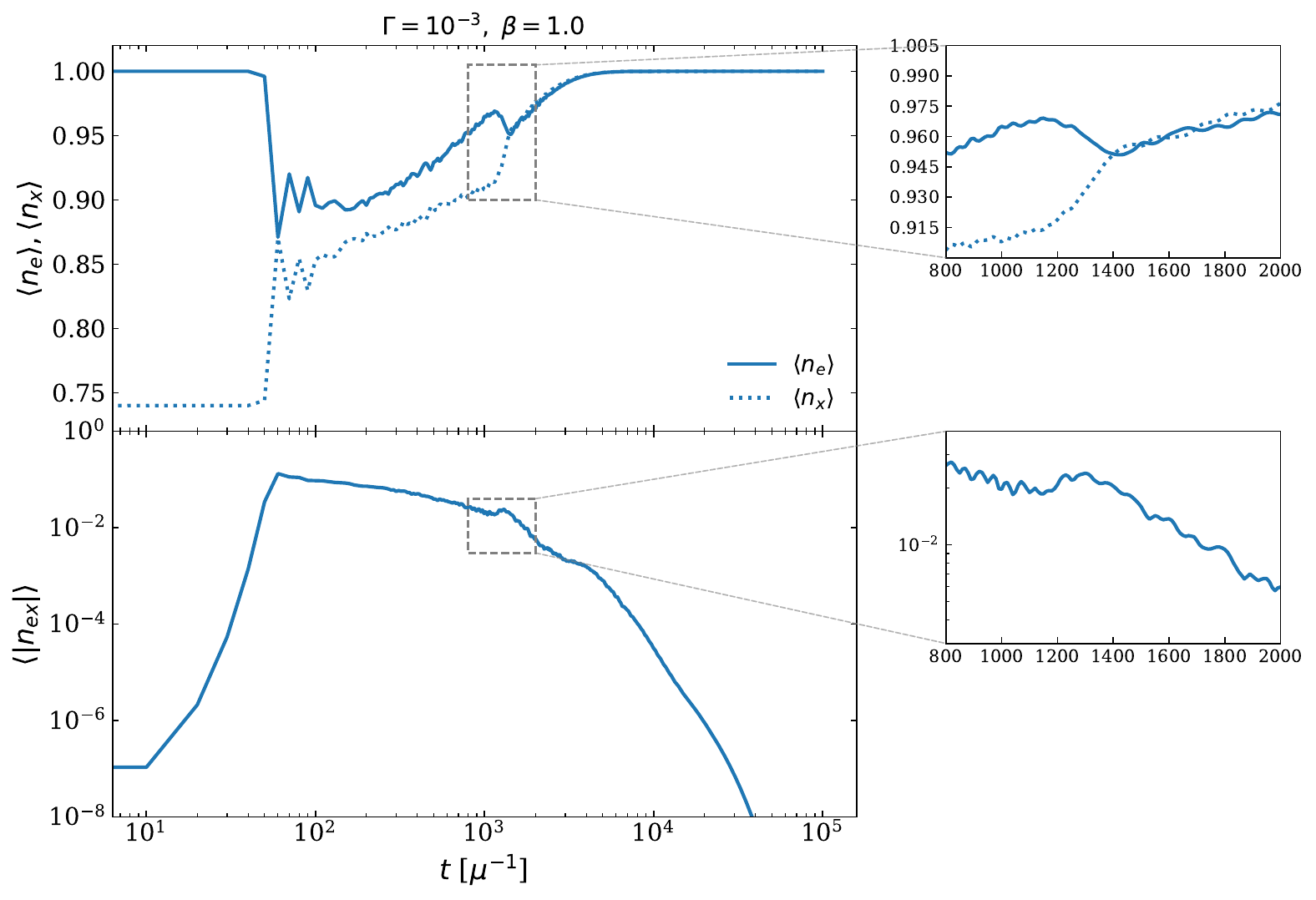}
    \caption{Same as Fig.~\ref{fig:magnify_sym}, but for the asymmetric case with $(\Gamma, \beta) = (10^{-3}, 1.0)$. The right panels focus on $t \in [800, 2000]$.}
    \label{fig:magnify_asym}
\end{figure}

Instead, we shall show in this subsection that the collisional modification of the neutrino angular distributions changes the FFI eigenmode structure, which drives the sudden flavor mixing. As representative examples, we focus on the symmetric case with $(\Gamma, \beta) = (10^{-4}, 0.1)$ and the asymmetric case with $(\Gamma, \beta) = (10^{-3}, 1.0)$; the time evolution around the sudden mixing events for these two cases is displayed in Figs.~\ref{fig:magnify_sym} and \ref{fig:magnify_asym}, respectively. While the underlying mechanism is essentially the same, the details differ between the symmetric and asymmetric cases, so we discuss them separately below.

\subsubsection{Symmetric case}
\label{sec:sym_case}

\begin{figure*}
    \centering
    \includegraphics[width=1\linewidth]{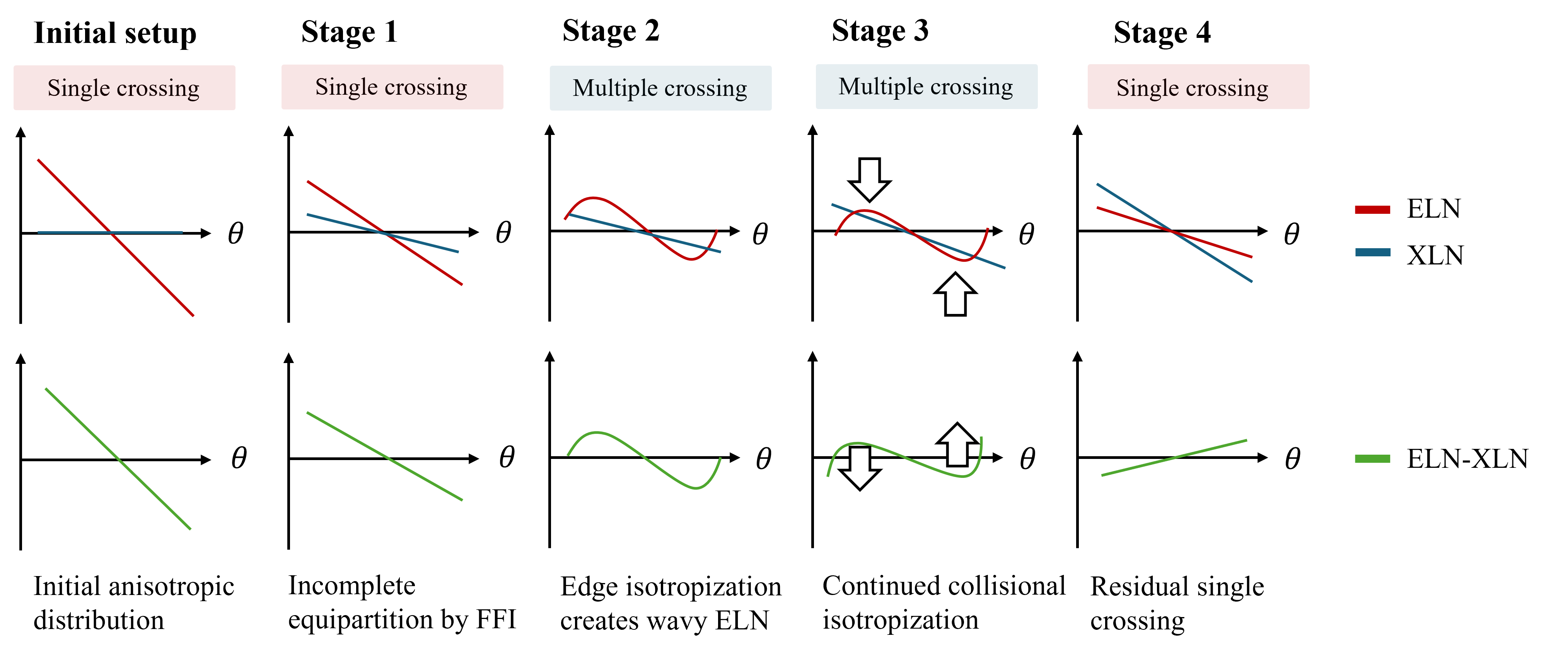}
    \caption{Schematic illustration of the four-stage temporal evolution of the ELN (red) and XLN (blue) angular distributions and the corresponding ELN$-$XLN (green) for the symmetric collision rate case. ``Initial setup'': the initial anisotropic distribution with a single crossing. ``Stage 1'' (incomplete equipartition by FFI): the FFI drives the system toward equipartition, but collisional damping prevents completion, leaving the ELN with a slightly steeper slope than the XLN. ``Stage 2'' (edge isotropization creates wavy ELN): the angle-dependent collisional relaxation isotropizes the forward and backward directions ($v \approx \pm 1$) faster, producing a wavy structure in the ELN and multiple zero-crossings in ELN$-$XLN. ``Stage 3'' (continued collisional isotropization): collisions further flatten the ELN, smoothing the wavy structure. ``Stage 4'' (residual single crossing): the system returns to a single-crossing configuration, shifting the dominant FFI eigenmode.}
    \label{fig:schematic}
\end{figure*}

We begin with introducing two useful quantities: the spatially and angularly integrated ELN and XLN number densities, defined as $N_{\rm ELN} \equiv \langle n_{\nu_e} \rangle - \langle n_{\bar{\nu}_e} \rangle$ and $N_{\rm XLN} \equiv \langle n_{\nu_x} \rangle - \langle n_{\bar{\nu}_x} \rangle$. As shown by \cite{Zaizen_2023}, in the absence of collisions, both $N_{\rm ELN}$ and $N_{\rm XLN}$ are conserved by the FFI under periodic boundary conditions. In the presence of collisions, their evolution is governed by
\begin{align}
 \frac{dN_{\rm ELN}}{dt}\bigg|_{\rm coll} &= (\Gamma - \bar{\Gamma})(n_{\nu_e}^{\rm eq} - \langle n_{\nu_e} \rangle) - \bar{\Gamma}\, N_{\rm ELN}, \label{eq:NELN} \\
 \frac{dN_{\rm XLN}}{dt}\bigg|_{\rm coll} &= 0, \label{eq:NXLN}
\end{align}
where $n_{\nu_e}^{\rm eq} = n_{\bar{\nu}_e}^{\rm eq}$ holds in our setup. The vanishing of $dN_{\rm XLN}/dt|_{\rm coll}$ follows directly from $\Gamma_x = \bar{\Gamma}_x = 0$. For $N_{\rm ELN}$, it also remains zero at all times in the symmetric collision rate case ($\Gamma = \bar{\Gamma}$) since $N_{\rm ELN}$ is set to be zero at $t=0$. Thus, both $N_{\rm ELN}$ and $N_{\rm XLN}$ are conserved throughout the evolution in the symmetric collision rate case.

\begin{figure}
    \centering
    \includegraphics[width=1\linewidth]{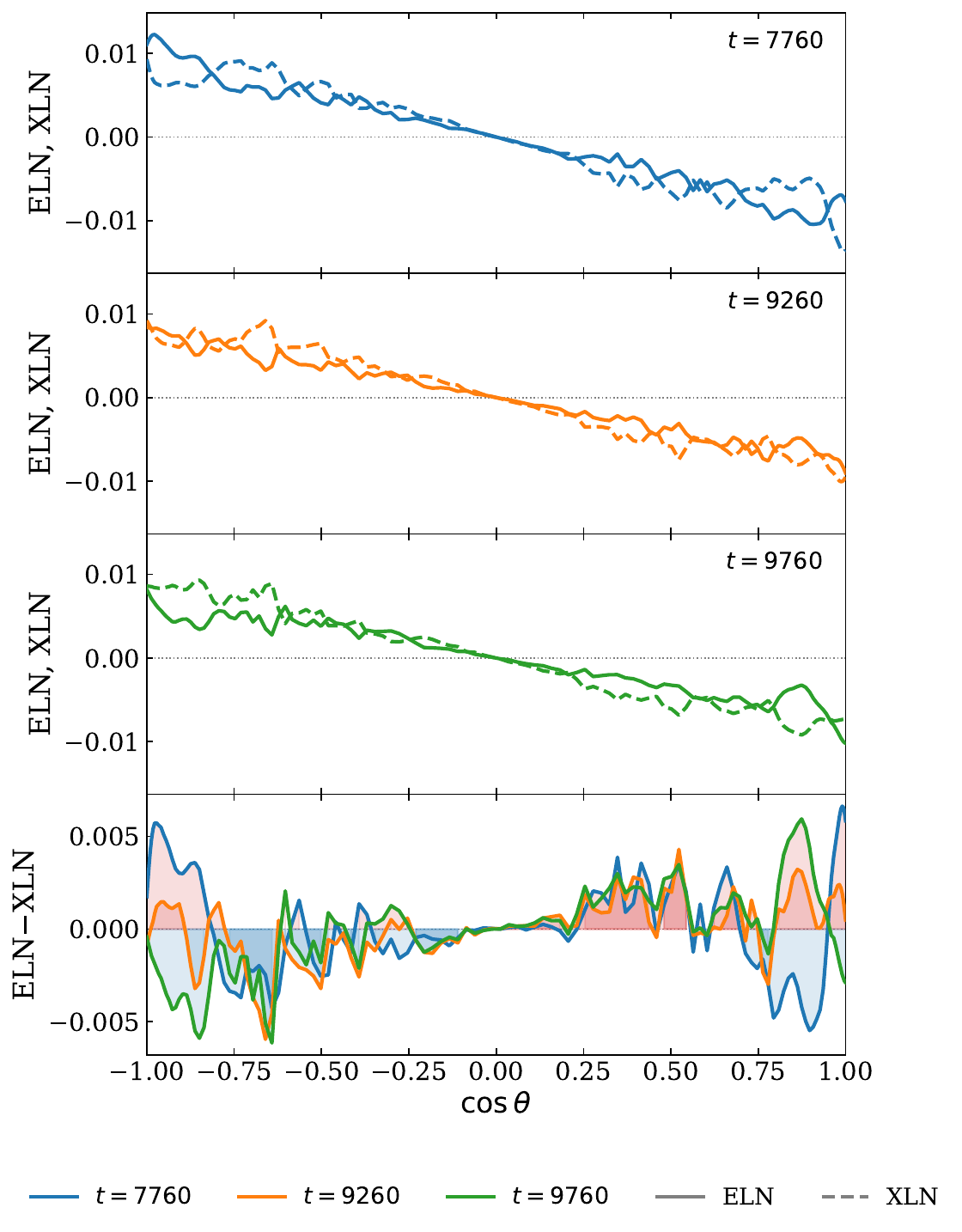}
    \caption{Spatially-averaged angular distributions of ELN (solid), XLN (dashed), and ELN$-$XLN (bottom panel) for the symmetric case with $(\Gamma, \beta) = (10^{-4}, 0.1)$. The distributions are shown at $t=7760$ (blue), $t=9260$ (orange), and $t=9760$ (green), corresponding to Stages 2, 3, and 4 in Fig.~\ref{fig:schematic}, respectively. The red (blue) shaded region in the bottom panel indicates where ELN$-$XLN is positive (negative).}
    \label{fig:angular_sym}
\end{figure}

The general picture for the symmetric case can be understood in four stages, as illustrated schematically in Fig.~\ref{fig:schematic}. Although the temporal fluctuations obscure some characteristic features of each state, the qualitative trends remain consistent with our simulations (see Fig.~\ref{fig:angular_sym}), lending direct support to our interpretation.

\begin{figure*}
    \centering
    \includegraphics[width=1\linewidth]{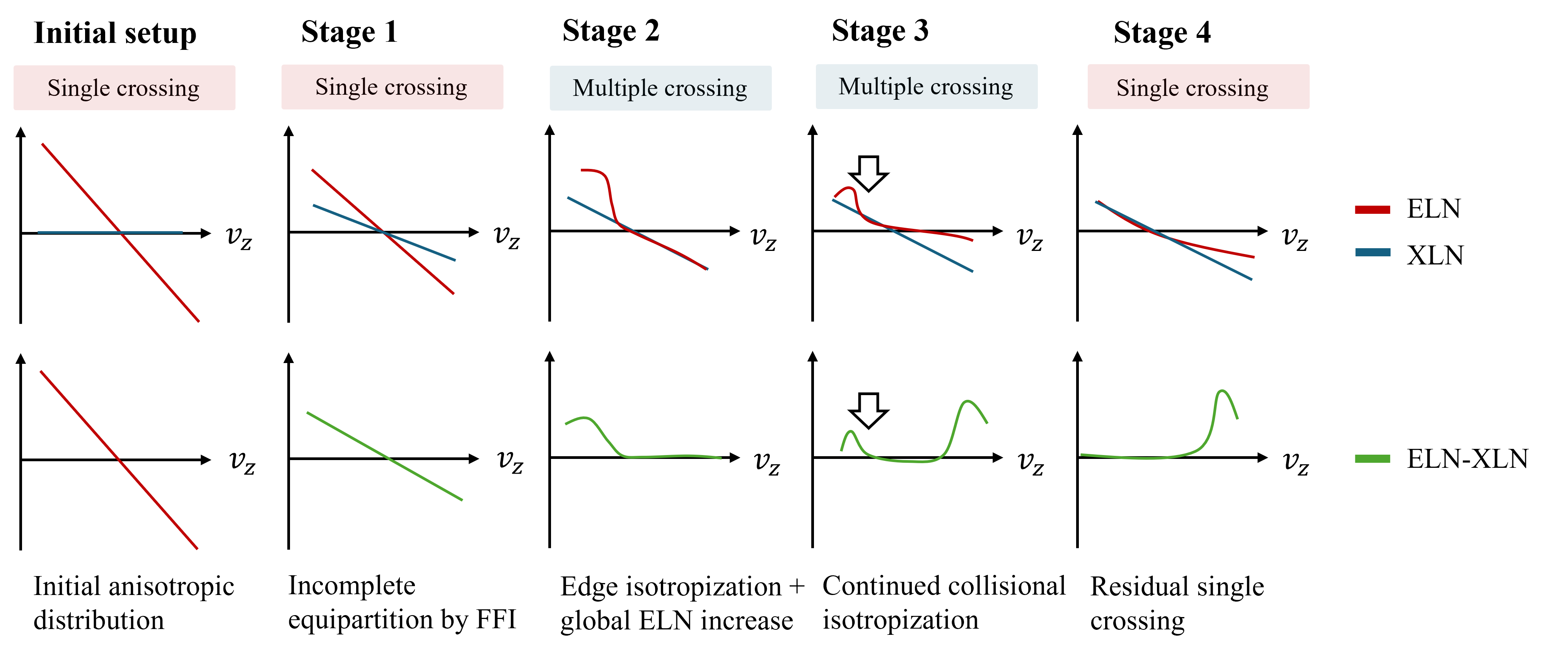}
    \caption{Same as Fig.~\ref{fig:schematic}, but for the asymmetric collision rate case ($\Gamma > \bar{\Gamma}$). The same four-stage sequence applies, but the non-conservation of $N_{\rm ELN}$ (Eq.~(\ref{eq:NELN})) introduces additional features. ``Stage 1'' (incomplete equipartition by FFI): same as the symmetric case. ``Stage 2'' (edge isotropization + global ELN increase): the backward direction develops a dominant positive peak in ELN$-$XLN, while the forward direction remains close to zero. ``Stage 3'' (continued collisional isotropization): the backward peak shrinks as $\bar{\nu}_e$ thermalization reduces the ELN there, while the forward peak grows. ``Stage 4'' (residual single crossing): the backward peak disappears and the system returns to a single-crossing configuration.}
    \label{fig:schematic_asym}
\end{figure*}

\textit{Stage 1: Incomplete equipartition by FFI.}--- The initial state has a single, large ELN$-$XLN angular crossing that triggers the primary FFI (see ``Initial setup'' in Fig.~\ref{fig:schematic}). Because the FFI timescale is shorter than that of collisions, the system initially evolves toward flavor equipartition via rapid flavor conversion across all angular directions. However, collisional damping inhibits complete equipartition, which can also be seen at $t \sim 10^3$ in Fig.~\ref{fig:magnify_sym}, where $\langle n_{\nu_e} \rangle$ is higher than $\langle n_{\nu_x} \rangle$. As a result, the ELN angular distribution retains a slightly steeper slope than the XLN, leaving behind a residual angular crossing (see ``Stage 1'' in Fig.~\ref{fig:schematic}).

\textit{Stage 2: Edge isotropization creates wavy ELN.}--- Subsequently, the collision terms $\Gamma(\rho_{ee}^{\rm eq} - \rho_{ee})$ and $\bar{\Gamma}(\bar{\rho}_{ee}^{\rm eq} - \bar{\rho}_{ee})$ act on $\nu_e$ and $\bar{\nu}_e$, respectively, driving their angular distributions toward the isotropic equilibrium \footnote{We note that the equilibrium ELN is zero in the current setup, since $\rho_{ee}^{\rm eq} = \bar{\rho}_{ee}^{\rm eq}$.}. Because the deviation from equilibrium is largest at the forward and backward directions ($v_z \approx \pm 1$), these angular bins relax more rapidly than those near the center ($\cos{\theta}=0$), producing a wavy structure in the ELN angular distributions (see ``Stage 2'' in Fig.~\ref{fig:schematic}). In contrast, $\nu_x$ and $\bar{\nu}_x$ are not directly affected by collisions ($\Gamma_x = 0$) and can only change through flavor conversion, so the XLN retains a nearly linear profile in their angular distributions. The combination of the wavy ELN and the nearly linear XLN gives rise to multiple zero-crossings in the ELN$-$XLN distribution.

It should be noted that the presence of multiple angular crossings caused by the wavy ELN angular distribution weakens the flavor instability (or increases the time scale of FFI); see, e.g., \cite{Bhattacharyya_2021}. As a result, angular crossings cannot be eliminated by FFI alone, since FFI is hindered by collisional damping in this regime, whereas collisions play a pivotal role in the subsequent quasi-steady evolution (see below).

\textit{Stage 3: Continued collisional isotropization.}--- As collisions continue to act, the ELN is progressively isotropized, and the wavy structure begins to smooth out (see ``Stage 3'' in Fig.~\ref{fig:schematic}). As a result, the multiple zero-crossings are gradually erased as the ELN distribution flattens.

\textit{Stage 4: Residual single crossing.}--- Collisional isotropization eventually drives the system back into a configuration with a single angular crossing (see ``Stage 4'' in Fig.~\ref{fig:schematic}). We note that FFI cannot change the sign of ELN-XLN at a local angular position \cite{Zaizen_2023}, implying that such a transition cannot be achieved by FFI alone.

Due to the qualitative change of angular distributions of ELN-XLN (transition from multiple angular crossings to single one), the eigenmodes of FFI are substantially modified, and the associated growth rate of FFI becomes larger than that in the regime with multiple angular crossings. Consequently, FFI overcomes the collisional damping, leading to the rapid flavor conversion and driving the system toward near flavor equipartition. At later times, the system evolves quasi-steadily toward its asymptotic state while preserving near flavor equipartition.

\subsubsection{Asymmetric case}
\label{sec:asym_case}

\begin{figure}
    \centering
    \includegraphics[width=1\linewidth]{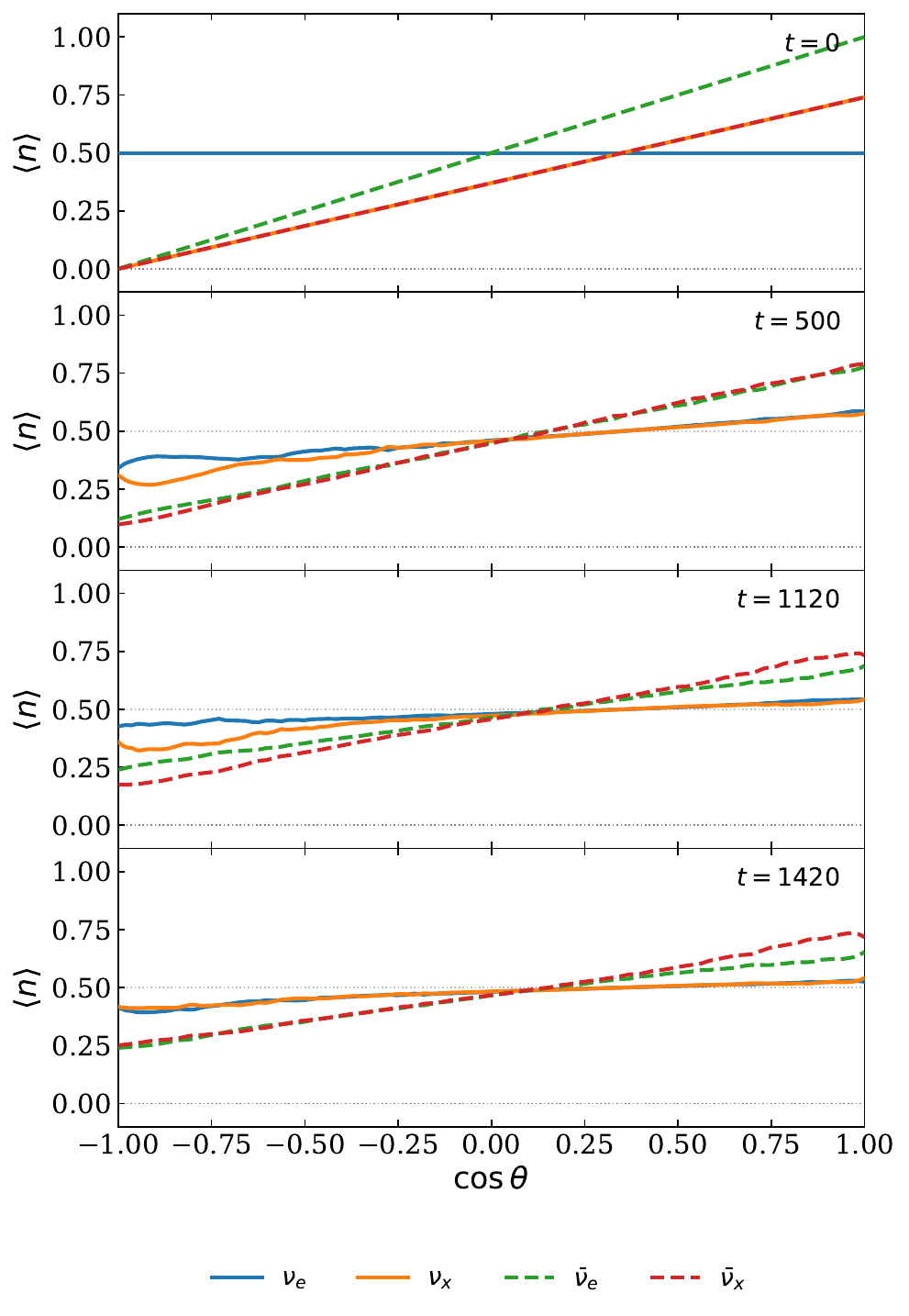}
    \caption{Spatially-averaged angular distributions of $\nu_e$ (blue solid), $\nu_x$ (orange solid), $\bar{\nu}_e$ (green dashed), and $\bar{\nu}_x$ (red dashed) for the asymmetric case with $(\Gamma, \beta) = (10^{-3}, 1.0)$ at four representative times: $t=0$ (initial setup), $t=500$ (stage 2), $t=1120$ (stage 3), and $t=1420$ (stage 4).}
    \label{fig:species_angular}
\end{figure}

In the asymmetric case, the time evolution of the system is broadly similar to that of the symmetric case, but the interplay between FFI and collisions becomes somewhat more complex due to the non-conservation of ELN number density (see below for more details). Here, we delve into the physical processes by focusing on the asymmetric collision rate model with $(\Gamma, \beta) = (10^{-3}, 1.0)$ (see also Fig.~\ref{fig:magnify_asym}).

Unlike the symmetric case where $N_{\rm ELN}$ remains zero (Sec.~\ref{sec:sym_case}), the asymmetric collision rates ($\Gamma > \bar{\Gamma}$) break this conservation, which can be understood through Eq.~(\ref{eq:NELN}). Since $n_{\nu_e}$ is depleted by FFI, leading to $\langle n_{\nu_e} \rangle < n_{\nu_e}^{\rm eq}$ throughout the evolution, the first term in the right-hand side of Eq.~(\ref{eq:NELN}) is always positive. Meanwhile, the second term is initially negligible since $N_{\rm ELN} \sim 0$, indicating that $N_{\rm ELN}$ becomes positive at $t>0$.

We note, however, that $N_{\rm ELN}$ does not monotonically increase with time in the entire phase. Instead, it decreases during the latter phase of evolution. As collisions drive $\nu_e$ towards equilibrium, the first term in the right-hand side of Eq.~(\ref{eq:NELN}) diminishes over time. In contrast, the second term grows in importance due to $N_{\rm ELN} >0$, and eventually dominates. As a result, $dN_{\rm ELN}/dt$ turns negative, causing $N_{\rm ELN}$ to decrease and asymptotically approach zero. This non-monotonic evolution of $N_{\rm ELN}$ is qualitatively distinct from the behavior observed in the symmetric case. It should be noted that $N_{\rm XLN}$ remains zero because $\Gamma_x = \bar{\Gamma}_x = 0$; thus this quantity is conserved even in the asymmetric case.

In Fig.~\ref{fig:schematic_asym}, we schematically illustrate the time evolution of ELN, XLN, and ELN-XLN angular distributions, analogous to Fig.~\ref{fig:schematic}. Although the overall evolution resembles that in the symmetric case, some distinct features emerge in the ELN angular distributions, particularly around $v_z \approx \pm 1$. Around $v_z \approx 1$, flavor equipartition is nearly achieved when the ELN-XLN angular crossings are almost diminished (stage 1). This can be attributed to the fact that ELN-XLN at $v_z \approx 1$ is negative (see ``Initial setup'' in Fig.~\ref{fig:schematic_asym}), while its angular integrated quantity ($N_{\rm ELN} - N_{\rm XLN}$) is positive at $t>0$, as discussed above. Under such conditions, angular regions with negative ELN-XLN (at $v_z \sim 1$) undergo strong flavor conversion \cite{Zaizen_2023}, driving neutrinos and antineutrinos toward flavor equipartition.

In contrast, flavor conversions in the angular region of $v_z \approx -1$ are less vigorous than those at $v_z \approx 1$, resulting in incomplete flavor equipartition. We can see this feature more clearly in Fig.~\ref{fig:species_angular}, which portrays species-dependent neutrino angular distributions at representative snapshots. At $t=500$ (the second panel from top), both neutrinos and antineutrinos achieve flavor equipartition for $v_z \gtrsim 0$, while deviations clearly remain for $v_z \lesssim 0$, consistent with the interpretation above. This angular-dependent flavor conversion can explain why $\langle n_{\nu_e} \rangle$ is higher than $\langle n_{\nu_x} \rangle$ during this phase (see $t \sim 500$ in Fig.~\ref{fig:magnify_asym}).

It should be noted, however, that such angular structures cannot be sustained in the later stages due to the continued effects of collisions. As in the symmetric case, collisions drive the ELN toward equilibrium (ELN $\to$ 0), qualitatively altering the structure of the ELN-XLN angular distribution (see ``Stage 2'' in Fig.~\ref{fig:schematic_asym}). Around $v_z \approx 1$, the ELN-XLN increases and eventually becomes positive, while the positive feature of ELN-XLN around $v_z \approx -1$ diminishes. These trends are primarily driven by $\bar{\nu}_e$ absorption and emission, respectively, as can be seen by comparing the panels in Fig.~\ref{fig:species_angular}: $t=500$ and $t=1120$ (second and third panels from top).

One may wonder why $\bar{\nu}_e$ collisions dominate the evolution despite the condition of $\Gamma > \bar{\Gamma}$, which suggests stronger $\nu_e$ interactions. The key lies in the detailed balance in collisions. As shown in Fig.~\ref{fig:schematic_asym}, $\nu_e$ distributions are more isotropic than those of $\bar{\nu}_e$, and are already closer to equilibrium across all angles. As a result, the detailed balance is nearly achieved for $\nu_e$, suppressing collisional effects. We also note that, although the collision rates themselves are angle-independent, the collision term is proportional to the deviation from the isotropic equilibrium [see Eqs.~(\ref{eq:C0}) and (\ref{eq:Cvec})], which varies with angle. For $\bar{\nu}_e$, this deviation is largest around $v_z \approx -1$, so the $\bar{\nu}_e$ emission replenishes the population most efficiently in that direction, causing the ELN there to decrease most rapidly.

Around $v_z \approx 0$, on the other hand, ELN and XLN are nearly equal to each other, and temporal variations generate multiple crossings in the ELN-XLN angular distribution. This indicates that the system lies near the edge of instability for FFI, and that flavor conversions remain active, albeit less vigorous, at this stage (see ``Stage 2'' in Fig.~\ref{fig:schematic_asym}).

\begin{figure}
    \centering
    \includegraphics[width=1\linewidth]{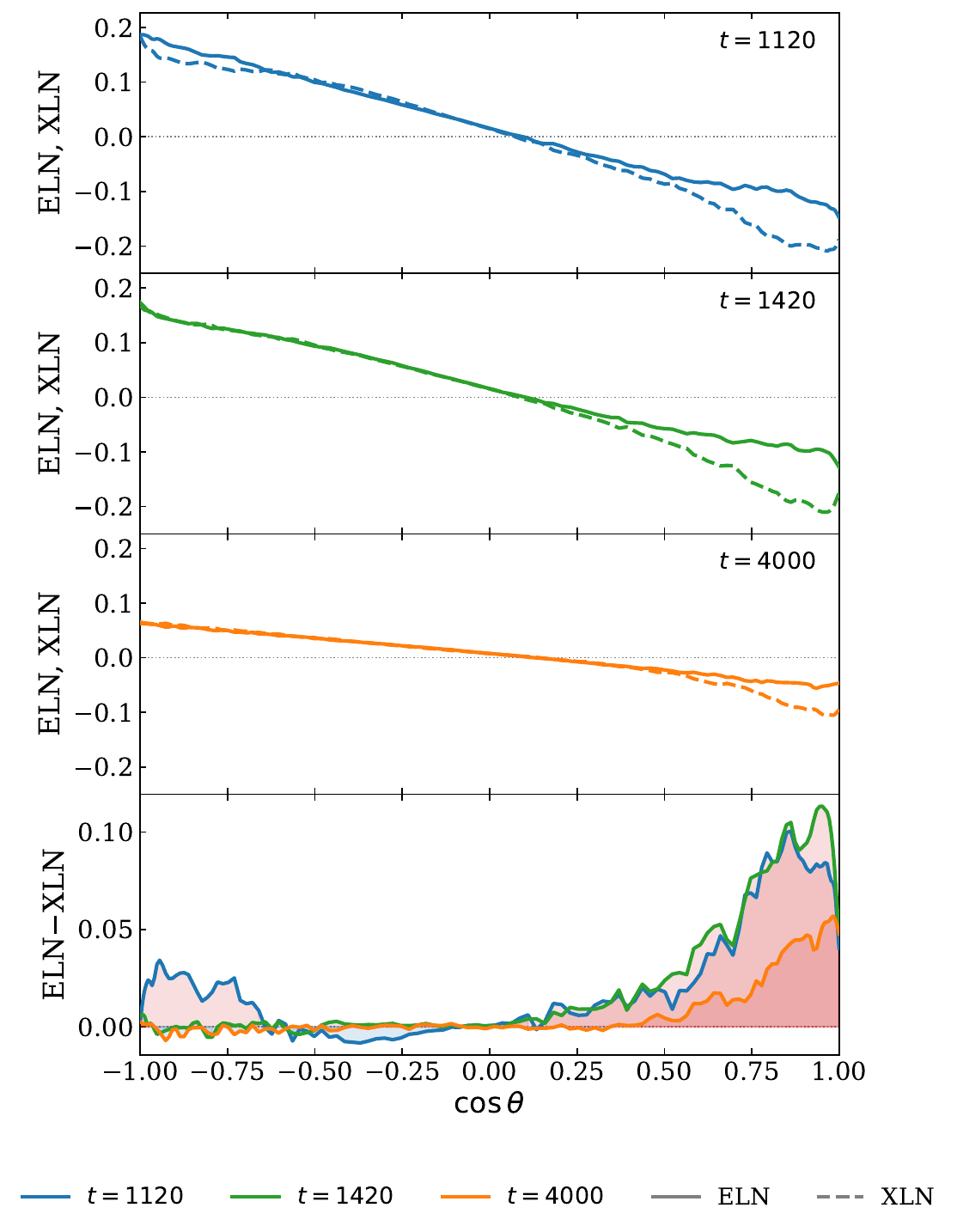}
    \caption{Spatially-averaged angular distributions of ELN (solid), XLN (dashed), and ELN$-$XLN (bottom panel) for the asymmetric case with $(\Gamma, \beta) = (10^{-3}, 1.0)$. The distributions are shown at $t=1120$ (blue), $t=1420$ (green), and $t=4000$ (orange). The red (blue) shaded region in the bottom panel indicates where ELN$-$XLN is positive (negative).}
    \label{fig:angular_asym}
\end{figure}

When the $\bar{\nu}_e$ emission at $v_z \approx -1$ drives ELN down to XLN, a single angular crossing forms in the ELN-XLN distribution, triggering a strong FFI. For the asymmetric model with $(\Gamma, \beta) = (10^{-3}, 1.0)$, this occurs around $t \sim 1270$ (corresponding to a time between the $t=1120$ and $t=1420$ panels shown in Fig.~\ref{fig:angular_asym}). In the top panel of Fig.~\ref{fig:angular_asym}, we display the ELN and XLN angular distributions at the time slightly before ELN crosses XLN at $v_z \approx -1$. As already pointed out above, FFI primarily develops in regions where ELN-XLN is negative (since $N_{\rm ELN} - N_{\rm XLN}$ is positive), implying that both neutrinos and antineutrinos evolve toward flavor equipartition at $v_z \approx -1$ (see ``Stage 3'' in Fig.~\ref{fig:schematic_asym}). Such a rapid phase transition occurs during the time of $1120 \lesssim t \lesssim 1420$, as evidenced by the disappearance of the positive ELN-XLN peak at $v_z \approx -1$ in Fig.~\ref{fig:angular_asym}.

An important remark concerns the impact of this late-time FFI. At first glance, Fig.~\ref{fig:angular_asym} might suggest that the impact is minor, since the changes in the ELN-XLN angular distribution are small. This interpretation is, however, misleading. A comparison of the bottom two panels in Fig.~\ref{fig:species_angular} ($t=1120$ and $1420$) shows that $\nu_e$ ($\bar{\nu}_e$) and $\nu_x$ ($\bar{\nu}_x$) are remarkably different at $t=1120$ around $v_z \approx -1$ (before the occurrence of late FFI), but become nearly identical at $t=1420$ (after the strong FFI episode). This trend supports the claim in \cite{Zaizen_2023} that the depth of ELN-XLN angular crossing does not reflect the impact of FFI (see Sec. III.C therein). In fact, this late-phase FFI is responsible for the rapid convergence of $\langle n_{\nu_e} \rangle$ and $\langle n_{\nu_x} \rangle$ displayed in the zoom-in panel of Fig.~\ref{fig:magnify_asym}. Its bottom panel also shows a clear increase of flavor coherence, indicating active FFI.

In the subsequent phase, the system undergoes quasi-steady evolution with the edge of instability. During this phase, FFI maintains flavor equipartition at $v_z \lesssim 0$, since collisions (or $\bar{\nu}_e$ emission) make ELN-XLN negative there. It is also important to note that $N_{\rm XLN}$ is a conserved quantity, implying that sustaining flavor equipartition at $v_z \lesssim 0$ requires adjustments of XLN in the $v_z \gtrsim 0$ regions. However, as $\nu_x$ is already in equilibrium with $\nu_e$ at $v_z \gtrsim 0$, $\nu_e \leftrightarrow \nu_x$ transitions cannot change XLN. Instead, the $\bar{\nu}_x$ distribution adjusts to conserve $N_{\rm XLN}$, reducing its population at $v_z \gtrsim 0$. We also note that the positive excess of $\bar{\nu}_e$ from their equilibrium state at $v_z \gtrsim 0$ is reduced through absorption. Through these processes, both $\bar{\nu}_x$ and $\bar{\nu}_e$ approach the equilibrium state, ultimately leading to flavor equipartition (and equilibrium state) across the entire angular domain. This gradual damping of the forward peak is clearly visible in the bottom panel of Fig.~\ref{fig:angular_asym}: by $t=4000$ (orange line), the positive ELN$-$XLN peak at $v_z \approx 1$ has substantially decreased in amplitude compared to its value at $t=1420$ (green line). This provides evidence that the above processes steadily drive the system toward the asymptotic equilibrium state.

\subsection{\label{sec:analysis_competition}Intermittent mixing with FFI, CFI, and collisions for asymmetric case}

\begin{figure}
    \centering
    \includegraphics[width=1\linewidth]{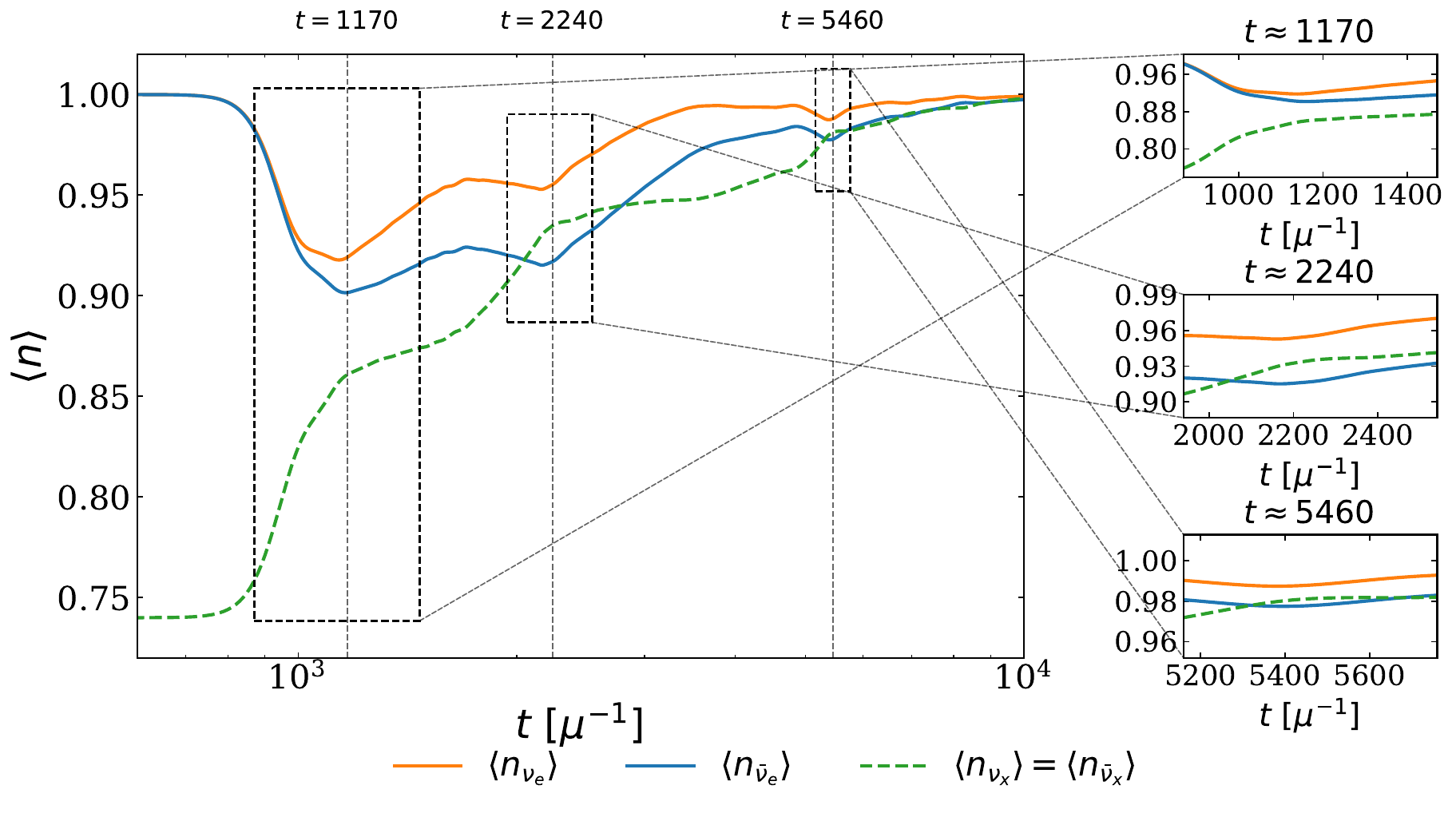}
    \caption{Time evolution of the spatially-averaged angular moments for the asymmetric intermediate model ($\beta = 0.1$, $\Gamma = 10^{-3}$, and $\bar{\Gamma} = 0.5\Gamma$). The main left panel shows $\langle n_{\nu_e} \rangle$ (solid orange), $\langle n_{\bar{\nu}_e} \rangle$ (solid blue), and $\langle n_{\nu_x} \rangle = \langle n_{\bar{\nu}_x} \rangle$ (dashed green). The three right panels provide magnified views of the mixing events at $t \approx 1170$ (top), $t \approx 2240$ (middle), and $t \approx 5460$ (bottom). Vertical dashed lines in the main panel mark these three times.}
    \label{fig:magnify_mixings}
\end{figure}

Here, we discuss another intriguing phenomenon appearing in the asymmetric model with intermediate crossing depth ($\beta = 0.1$, $\Gamma = 10^{-3}$, and $\bar{\Gamma} = 0.5\Gamma$), where the growth timescales of the FFI and CFI become comparable to the collision timescale ($\tau_{\rm FFI} \approx \tau_{\rm CFI} \lesssim \tau_{\rm col}$). In this regime, flavor conversion occurs intermittently, reflecting the competing influences of FFI, CFI, and collisional thermalization (see also orange lines in Fig.~\ref{fig:asymmetric_high_gamma}). In this model, we identify three key phases, during which $\langle n_{\nu_e} \rangle$ ($\langle n_{\bar{\nu}_e} \rangle$) and $\langle n_{\nu_x} \rangle$ ($\langle n_{\bar{\nu}_x} \rangle$) rapidly converge to each other at $t \approx 1170$, $2240$, and $5460$, as illustrated in Fig.~\ref{fig:magnify_mixings}. As we shall discuss below, the first and second events are predominantly driven by FFI, while the third one is solely driven by CFI. Collisional thermalization also plays an important role during the transition phases. In the following, we describe these dynamical processes in chronological order.

\subsubsection{First mixing event by FFI}

The first mixing event at $t \approx 1170$ is primarily driven by FFI. In this model, the FFI timescale is somewhat shorter than the resonance-like CFI, which can be seen by comparing with green and orange curves, where the green curve corresponds to a model with dominant resonance-like CFI in Fig.~\ref{fig:asymmetric_high_gamma}. On the other hand, the growth of the FFI is suppressed by collisions, as indicated by the comparison between the gray and orange lines in Fig.~\ref{fig:asymmetric_high_gamma} (see also the discussions in Sec.~\ref{sec:overall_features}). Such collisional damping affects not only the linear phase but also the nonlinear phase; in fact, the degree of flavor conversion during the first FFI episode is quantitatively smaller than in the collisionless case. This relatively weaker flavor mixing, compared to other models, leads to more complex subsequent dynamics (see below).

It is also worth commenting on the CFI during this phase. The first FFI tends to drive $G$ [see Eq.~(\ref{eq:CFI_primitive}) for the definition of $G$] towards zero through $\nu_e \leftrightarrow \nu_x$ and $\bar{\nu}_e \leftrightarrow \bar{\nu}_x$ transitions, thereby pushing the system toward a stable CFI configuration. This can be seen in the bottom panel of Fig.~\ref{fig:cfi_growthrate}, which shows that $G$ drops sharply (from $G \sim 0.26$ to $G \sim 0.04$) during the first mixing. We also note that the condition of resonance-like CFI is satisfied at $t=0$ [since $A=0$; see also Eqs.~(\ref{eq:CFI_primitive}) and (\ref{eq:CFIgrowthrate_isopre})], but this condition is subsequently violated by the onset of FFI. Although $A$ is not directly affected by flavor conversions [see also Eq.~(\ref{eq:NELN})\footnote{From the definitions, $A = (N_{\rm ELN} - N_{\rm XLN})/2$, and $N_{\rm XLN}$ remains zero throughout the evolution in our models, so that $A = N_{\rm ELN}/2$.}], it increases shortly after the onset of FFI (see the bottom panel of Fig.~\ref{fig:cfi_growthrate}). This occurs because $\nu_e$ and $\bar{\nu}_e$ deviate from the equilibrium, driving an increase in $A$. Consequently, the resonance condition is no longer satisfied, indicating that any subsequent CFI, if it occurs, must be of the non-resonant type.

\subsubsection{Transition phase between the first and second mixing episodes: $1170 \lesssim t \lesssim 2240$}

\begin{figure}
    \centering
    \includegraphics[width=1\linewidth]{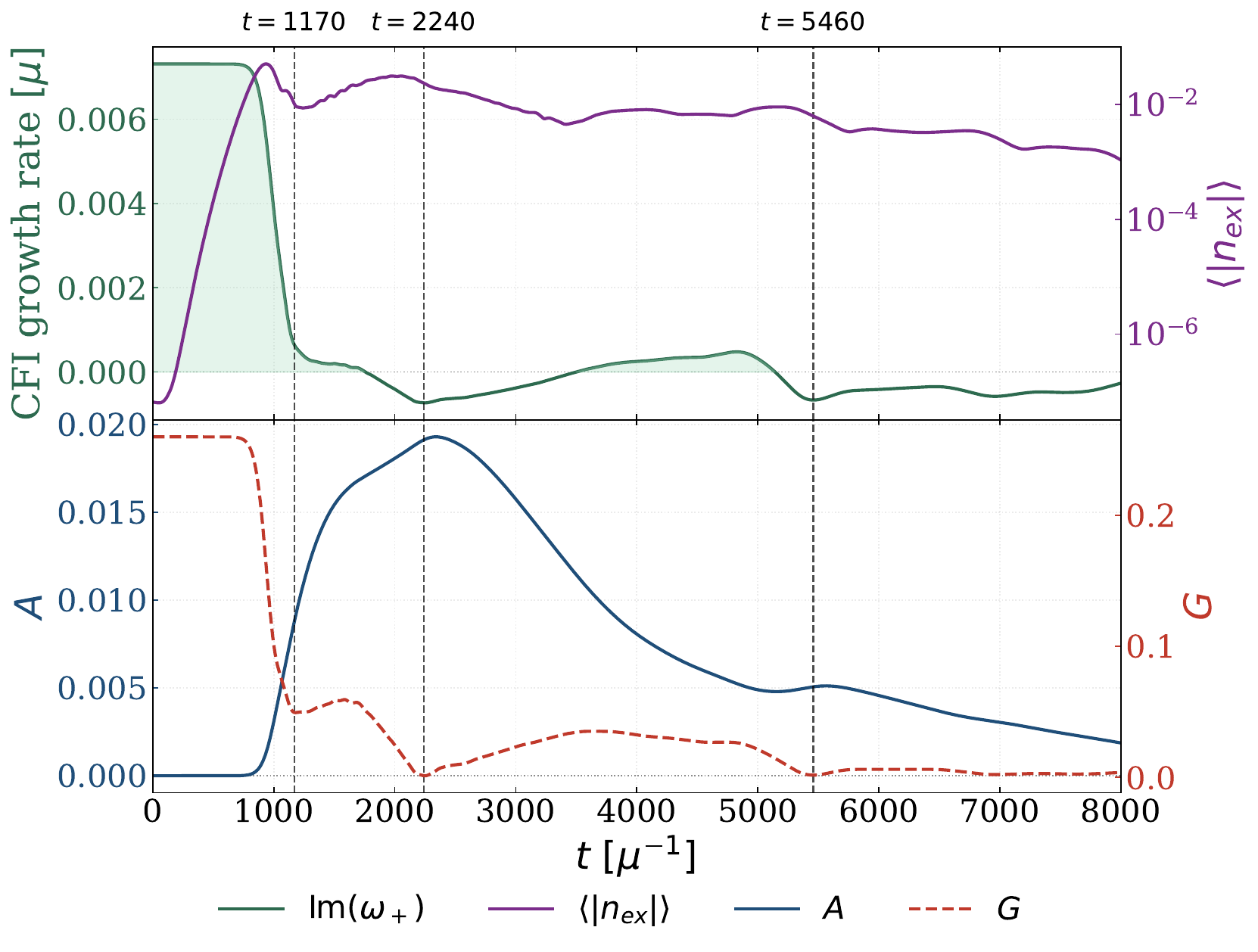}
    \caption{Time evolution of quantities relevant to the CFI for the asymmetric intermediate model ($\beta = 0.1$, $\Gamma = 10^{-3}$, $\bar{\Gamma} = 0.5\Gamma$). The top panel shows the CFI growth rate $\mathrm{Im}(\omega_+)$ computed from Eqs.~(\ref{eq:dispersions})--(\ref{eq:CFIgrowthrate_isopre}) (solid green, left axis, with the positive region shaded) together with the spatially-averaged off-diagonal amplitude $\langle |n_{\rm ex}| \rangle$ (solid purple, right axis, log scale). The bottom panel shows the ELN--XLN number density $A$ (solid blue, left axis) and the population asymmetry $G$ (dashed red, right axis). Vertical dashed lines mark the three mixing events at $t \approx 1170$, $2240$, and $5460$.}
    \label{fig:cfi_growthrate}
\end{figure}

\begin{figure*}
    \centering
    \includegraphics[width=1\linewidth]{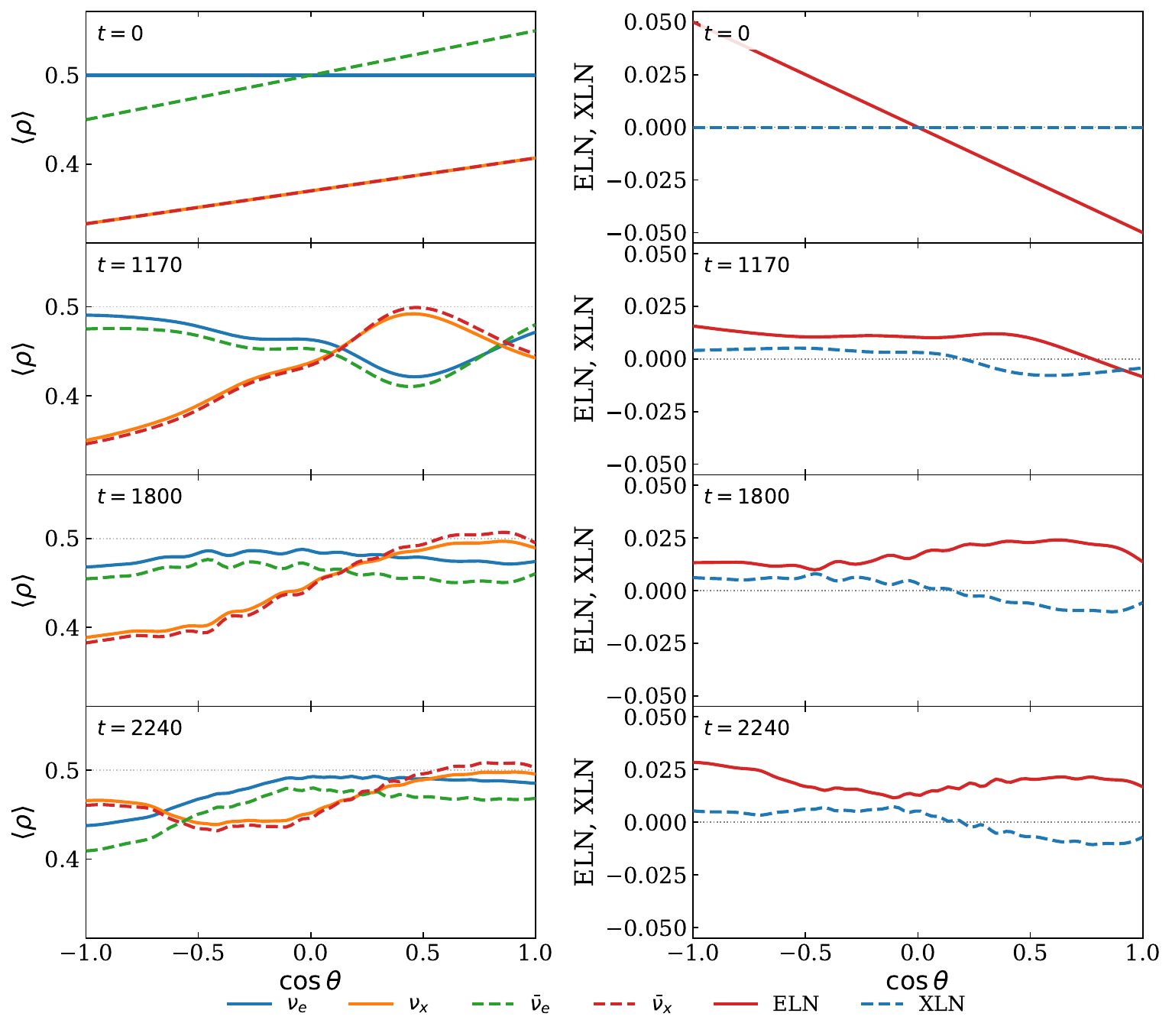}
    \caption{Spatially-averaged angular distributions of the four species (left column) and of ELN (red solid) and XLN (blue dashed) (right column) for the asymmetric intermediate model ($\beta = 0.1$, $\Gamma = 10^{-3}$) at four representative times: $t=0$ (initial setup), $t=1170$ (first mixing), $t=1800$ (during the inter-mixing interval), and $t=2240$ (second mixing).}
    \label{fig:species_angular_asym_int}
\end{figure*}

In the short time window $1170 \lesssim t \lesssim 1500$ (immediately following the first FFI), the growth rate of CFI ($\mathrm{Im}(\omega_+)$), which is displayed in the top panel of Fig.~\ref{fig:cfi_growthrate}, remains positive. This residual CFI may contribute to the growth of $\langle |n_{\rm ex}| \rangle$ during this phase. However, $A$ continues to grow and reaches $A \sim 0.015$ at $t \sim 1500$, while $G$ remains nearly constant at $G \sim 0.04$. As a result, $G\alpha/|A|$ decreases, driving $\mathrm{Im}(\omega_{\pm})$ down to and eventually below zero, indicating that CFI subsides.

\begin{figure*}
    \centering
    \includegraphics[width=1\linewidth]{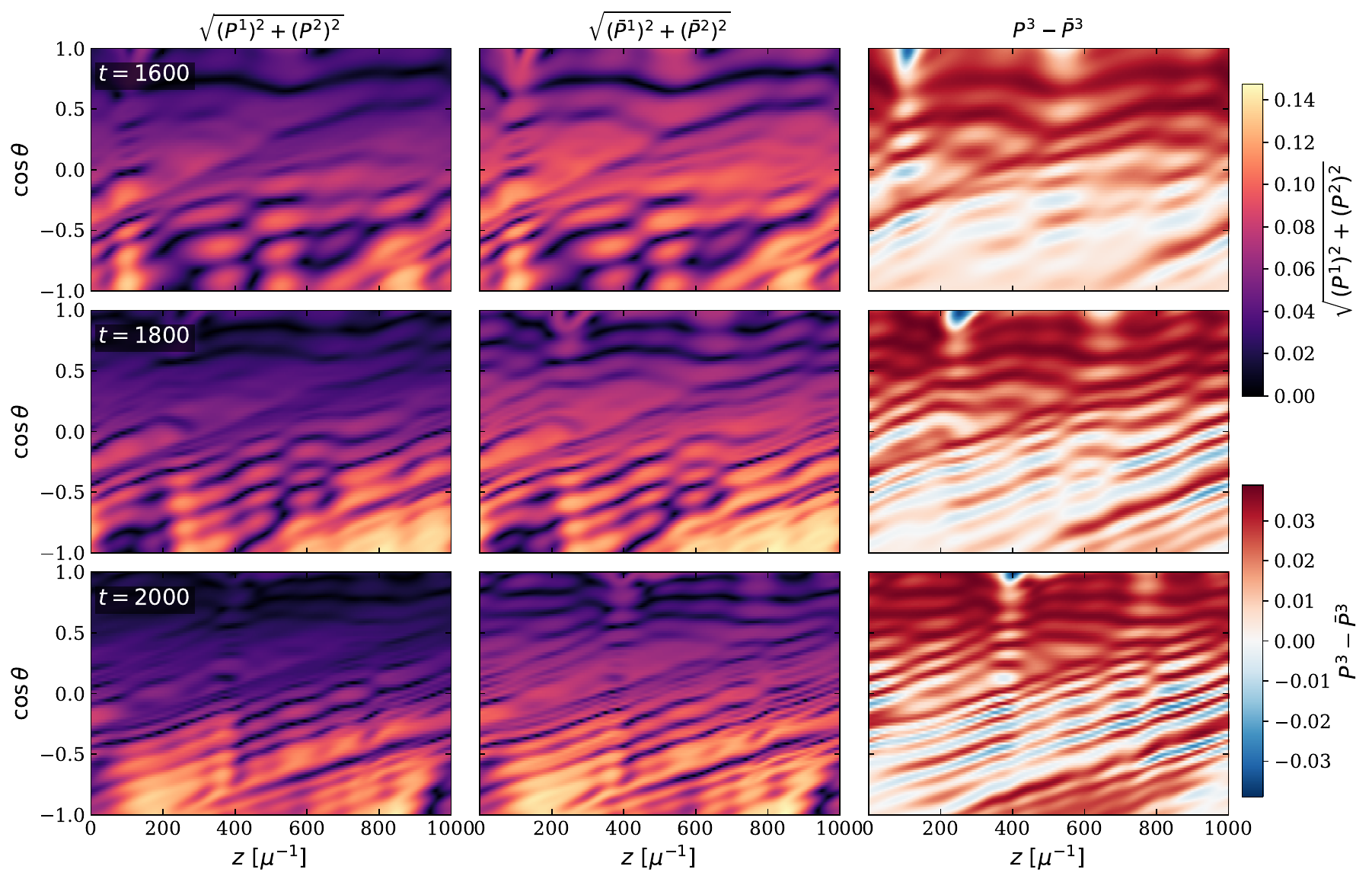}
    \caption{Spatial and angular distributions of the polarization-vector amplitudes $\sqrt{(P_1)^2 + (P_2)^2}$ for neutrinos (left column) and antineutrinos (middle column), and $P_3 - \bar{P}_3$ (right column), for the asymmetric intermediate model ($\beta = 0.1$, $\Gamma = 10^{-3}$) at three times preceding the second mixing event: $t=1600$ (top row), $t=1800$ (middle row), and $t=2000$ (bottom row). The flavor coherence grows preferentially at backward angles ($v_z \lesssim 0$), indicative of a local FFI.}
    \label{fig:phasespace_local}
\end{figure*}

Despite the absence of CFI in the interval $1500 \lesssim t \lesssim 2000$, $\langle |n_{\rm ex}| \rangle$ continues to grow [see the top panel of Fig.~\ref{fig:cfi_growthrate}], indicating that some flavor instability remains active during this phase. To clarify the origin of these flavor conversions, we examine the possibility of FFI. In Fig.~\ref{fig:species_angular_asym_int}, we show spatially-averaged angular distributions of the four neutrino species together with the ELN and XLN at four representative times. Interestingly, no remarkable ELN-XLN angular crossings are observed during the phase (see right panels). However, ELN and XLN are nearly equal to each other at the angular position of $v_z \sim -0.4$ at $t \sim 1800$, indicating that the system lies near the threshold of FFI.

Notably, FFI can occur locally even in such a globally metastable state. To verify this claim, we present the spatial and angular distributions of the polarization-vector amplitudes $\sqrt{(P_1)^2 + (P_2)^2}$ for $\nu$ and $\bar{\nu}$, together with $P_3 - \bar{P}_3$, at $t=1600$, $1800$, and $2000$ in Fig.~\ref{fig:phasespace_local}. At all three times, the flavor coherence $\sqrt{(P_1)^2 + (P_2)^2}$ grows preferentially at backward angles ($v_z \lesssim 0$), while the forward region remains nearly unperturbed. In the same backward region, a local $P_3 - \bar{P}_3$ (ELN--XLN) crossing is visible, suggesting that a local FFI is operating in the backward region. This feature is consistent with Fig.~\ref{fig:species_angular_asym_int}, showing that strong flavor conversions occur at $v_z <0$ during this phase (see the left panels for $t=1800$ and $2240$).

\subsubsection{Termination of the local FFI around $t \sim 2240$}

As shown in Figs.~\ref{fig:magnify_mixings}~and~\ref{fig:cfi_growthrate}, the local FFI subsides at $t \sim 2240$. At this time, the ELN--XLN angular distribution exhibits a widened gap at $v_z \lesssim 0$ (see bottom row of Fig.~\ref{fig:species_angular_asym_int}): the ELN has increased there, while the XLN remains nearly unchanged. This behavior can be understood as follows. Prior to the onset of local FFI, $\nu_e$ and $\bar{\nu}_e$ at $v_z \lesssim 0$ are close to their equilibrium states (see the left panel at $t=1170$ in Fig.~\ref{fig:species_angular_asym_int}), indicating that collisions play a subdominant role. However, once the local FFI shuffles neutrino flavors, both $\nu_e$ and $\bar{\nu}_e$ deviate from the equilibrium state (see left panels at $t=1800$ and $2240$ in Fig.~\ref{fig:species_angular_asym_int}). Because $\Gamma > \bar{\Gamma}$, collisions drive $\nu_e$ back towards the equilibrium more rapidly than $\bar{\nu}_e$, leading to an increase of ELN. When the ELN at $v_z \lesssim 0$ becomes sufficiently large to eliminate the local ELN-XLN angular crossings, the local FFI is consequently terminated.

\subsubsection{Collisional thermalization and transition to the non-resonant CFI ($t \gtrsim 2240$)}

In the subsequent phase ($t \gtrsim 2240$), collisions govern the time evolution of the system. As shown in the bottom panel of Fig.~\ref{fig:cfi_growthrate}, $G$ increases with time, while $A$ decreases until $t \sim 3500$. As a result, the system enters an unstable regime for non-resonant CFI (see the green line in the top panel of Fig.~\ref{fig:cfi_growthrate}). Consequently, the decline in flavor coherence ($|n_{\rm ex}|$) halts at $t \sim 3500$, after which it remains nearly constant until $t \sim 5000$ (see the purple line in the top panel of Fig.~\ref{fig:cfi_growthrate}).

It should be noted, however, that the non-resonant CFI is initially too weak to induce significant flavor conversions. In fact, $G$ remains nearly constant up to $t \sim 5000$, indicating that collisional thermalization competes with the non-resonant CFI during this interval. As time progresses, however, the efficiency of thermalization diminishes as the system approaches the detailed balance between emission and absorption. Meanwhile, the CFI growth rate increases with decreasing $A$. Consequently, the non-resonant CFI eventually overtakes the collisional thermalization at $t \sim 5000$, triggering strong flavor conversions (which corresponds to the origin of the third strong mixing event). After the flavor conversion saturates, the system enters a final quasi-steady phase and asymptotically approaches the universal flavor-equilibrated state described in Sec.~\ref{sec:overall_features}.

\section{\label{sec:discussion}Summary and Discussion}

Recent detailed investigations of flavor instabilities based on multi-dimensional CCSN simulations have revealed that spatial regions exhibiting resonance-like CFI overlap with those associated with FFI \cite{Akaho_2024}. This primarily arises because the condition for resonance-like CFI is satisfied in regions where the number densities of $\nu_e$ and $\bar{\nu}_e$ are nearly equal. In such environments, ELN angular crossings can naturally occur, since the angular distribution of $\bar{\nu}_e$ is, in general, more forward peaked than that of $\nu_e$. In this regime, however, the dynamics of flavor conversion in the non-linear phase remains highly uncertain. This uncertainty stems from the complex interplay among the FFI, CFI, and collisions, whose dynamics are expected to depend sensitively on neutrino properties (e.g., species dependent angular distributions) and collision rates. Motivated by these considerations, we performed spatially inhomogeneous numerical simulations of quantum kinetic neutrino transport that simultaneously incorporate the dynamics of FFI, CFI, and collisional relaxation by varying neutrino angular distributions and collision rates. Our main findings are summarized as follows:

\begin{itemize}
    \item \textit{Universal asymptotic state.} For all models with strong flavor conversions, the system converges to the same flavor-equilibrated state regardless of the intermediate dynamics. The only exception is the shallow-crossing case with symmetric collision rates (i.e., no CFI in this model). In this model, FFI occurs in the very early phase but fails to drive strong flavor conversions due to collisional damping (see Sec.~\ref{sec:overall_features} and Fig.~\ref{fig:summary} for more details).
    \item \textit{Diverse intermediate pathways.} Despite converging to a universal asymptotic state, the intermediate dynamics exhibit rich and model-dependent structures. Their overall behavior can be categorized by the hierarchy among $\tau_{\rm FFI}$, $\tau_{\rm CFI}$, and $\tau_{\rm col}$. We also find that the results differ significantly between symmetric and asymmetric collision rates, indicating that CFI drives more complex dynamics. See Secs.~\ref{sec:sym} and \ref{sec:asym} for more details.
    \item \textit{Collision-induced FFI eigenmode transition.} One of the intriguing phenomena observed in this study is that collisions substantially modify the neutrino angular distributions during the evolution and thereby trigger an FFI eigenmode transition. This reinvigorates flavor conversion long after the initial FFI saturates. Our result suggests that collisions can act not only as a damping mechanism for FFI but also as an active driver of subsequent FFI activity (Sec.~\ref{sec:analysis_modification}).
    \item \textit{Multi-stage mixing from FFI--CFI competition.} When the timescales of FFI and resonance-like CFI are comparable initially, which corresponds to the intermediate regime with asymmetric collision rates in our classification (see Fig.~\ref{fig:summary}), the system undergoes multiple phases of strong flavor conversion during its evolution. Transitions between these phases are also influenced by collisional thermalization; see Sec.~\ref{sec:analysis_competition} for further details.
\end{itemize}

While our results provide new insights into the complex dynamics of neutrino flavor conversion, our current numerical framework contains several limitations. First, due to limited computational resources, the collision rates employed in our simulations were artificially enhanced relative to those in actual CCSNe. However, since our study systematically categorizes the dynamics based on the relative balance among the characteristic timescales ($\tau_{\rm FFI}$, $\tau_{\rm CFI}$, $\tau_{\rm col}$) rather than their absolute values, we expect that the essential physical features are still captured. Second, the periodic boundary conditions employed in this study may be inappropriate for describing realistic CCSN environments. Addressing this issue requires global simulations of quantum kinetic neutrino transport. However, such simulations in regimes with high neutrino densities remain computationally intractable, even when employing attenuation prescriptions commonly used in the literature \cite{Nagakura_2022, Xiong_2023, Nagakura_2023, Nagakura_2023_PRL, Nagakura_2023_basic, Urquilla_2026, Zaizen_2026}. Third, our initial conditions do not satisfy the steady-state constraint that would be nearly achieved in realistic CCSN environments; see, e.g., the discussion in \cite{Johns_2022}. Addressing this issue also requires a self-consistent global framework, and the interplay between transport and flavor conversion would naturally determine the angular distributions, eliminating this artificial freedom. Finally, our simplified single-energy, two-flavor collision term does not capture the energy dependence of weak interaction processes or the role of the muon neutrino sector. Incorporating multi-energy bins, full scattering kernels, and a three-flavor treatment---which can introduce additional sources of CFI \cite{Jiabao_2024_Muon}---are important directions for future work.

Despite these limitations, our finding that the dynamic competition among FFI, CFI, and collisions tends to drive the system toward a universal flavor equilibrium provides an important clue for developing improved subgrid models for collective neutrino oscillations in regimes where FFI and CFI coexist. Importantly, our present study suggests that the resulting non-linear dynamics can differ significantly from those governed by any single flavor instability alone. We leave the development of subgrid models for such mixing modes (FFI and CFI), incorporating more realistic collisional effects, to future work.

\begin{acknowledgments} S.T. is supported by Grant-in-Aid for JSPS Fellows (26KJ0991), a JSR fellowship and the International Graduate Program of Innovation for Intelligent World. H.N. is supported by Grant-in-Aid for Scientific Research (23K03468) and the HPCI System Research Project (Project ID: hp250006, hp250226, hp250166, hp260058). M.Z. is supported by JSPS KAKENHI Grant Numbers JP24H02245 and JP25K17383. C.K. is supported by JSPS KAKENHI Grant Numbers JP24H02245 and JP25K17395. J.L. is supported by Grant-in-Aid for JSPS Fellows (26KJ2052). These simulations were performed using computational resources supported by JSPS KAKENHI Grant Number 24H00004.
\end{acknowledgments}

\bibliographystyle{apsrev4-2} \bibliography{ref}
\end{document}